\documentclass{style}
\usepackage{graphicx}
\pdfoutput=1
\begin{document}
\title{Interplay between one-particle and collective degrees of freedom in
nuclei} 

\author{ Ikuko Hamamoto\email{Ikuko.Hamamoto@MATFYS.LTH.SE} \\
\it Riken Nishina Center, Wako, Saitama 351-0198, Japan  \\ 
\it Division of Mathematical Physics, Lund Institute of Technology 
at the University of Lund, Lund, Sweden }   


\date{}

\maketitle

\begin{abstract}
Some developments of nuclear-structure physics uniquely related 
to Copenhagen School are sketched 
based on theoretical considerations versus experimental
findings and one-particle versus collective aspects.  Based on my personal 
overview I pick up the following topics;
(1) Study of vibration in terms of particle-vibration coupling; (2) One-particle
motion in deformed and rotating potentials, and yrast spectroscopy in high-spin 
physics; (3) Triaxial shape in nuclei: wobbling motion and chiral bands; 
(4) Nuclear structure of drip line nuclei: in particular, shell-structure (or
magic numbers) change and spherical or deformed halo phenomena; (5) shell
structure in oblate deformation.

\end{abstract}


\section{Introduction} 
I came to Copenhagen in September, 1967, holding one-year's fellowship from
the Nishina Foundation in Japan.  It was my first visit to foreign institutes,
where I had to understand physics discussed in English.  Indeed, it took 
several years for me to understand some physics which I heard in English.   
Everything was new and interesting to me, but among others, I remember 
clearly "experimental meeting at NBI (= Niels Bohr Institute)" at 11 o'clock 
on Monday morning, in which I learned how to talk with and study from 
experimentalists.  New experimental data presented 
by experimentalists coming from all over the world were of course exciting,
however, it was my strong impression that people were more carefully watching 
the questions and reactions by Aage Bohr and Ben Mottelson 
to those presentations.  In the sixties at NBI we could meet physicists from
both West (USA, Canada, Europe, Australia, etc.) and East (Soviet Union,
East Europe, China, etc.), and we became good friends for life 
after we spent our younger days together.  

My stay in Copenhagen, which at the beginning I intended for one year, became
for three years.  Though I once left Copenhagen in the summer of 1970, I came
back to Europe already in 1971 and to Copenhagen in 1973.  Since 1971 I helped
the completion of Bohr and Mottelson's book, NUCLEAR STRUCTURE vol. II, 
which was
finally published in 1975.  
The book is not at all just a textbook, but the entire volume is a big article,
which is full of their original ideas and deeply-going understanding of physics.  
During those years I learned an enormous amount of
physics from Bohr and Mottelson especially 
because I could directly talk with them and 
ask questions to them almost whenever I wanted.  It was the exclusively precious
and happy 
time and days in my life as a physicist, 
which I never forget, and I want to express 
my heartfelt thanks to Aage and Ben.      

In the present article I write my personal overview of five topics 
that have been centrally placed in the field of nuclear-structure physics 
in respective decades.  Those topics perhaps 
except for the most recently developed drip-line physics have been   
developed in the way either strictly followed from the ideas of 
Bohr and Mottelson 
or strongly influenced by their way of thinking physics, and
the topics are those, the study of which I myself have also eagerly worked in.  
In Sec. II particle-vibration coupling,
in Sec. III one-particle motion in deformed and rotating potentials, 
and yrast spectroscopy in high-spin physics, 
in Sec. IV triaxial shape of nuclei, 
in Sec. V nuclear structure as neutron-drip-line approaches, and in Sec. VI
nuclear shell-structure in oblate deformation are presented, 
while conclusions and discussions are given in Sec. VII.

\section{Particle-vibration coupling} 
In the self-consistent system such as nuclei one-particle motion and collective
phenomena are strongly related.
Elementary modes of excitations may be associated with excitations of individual
particles or they may represent collective vibrations of the density, shape, or
some other parameter that characterizes the equilibrium configuration.  
The vibrational motion in nuclei is so profoundly affected 
by the shell structure of
one-particle motion that it presents an excellent example of the interweaving of
one-particle and collective degrees of freedom \cite{BM75}.  

In the nuclear system the possibility of collective shape oscillations is
strongly suggested by the fact that the ground states of some nuclei are
described by densities and mean fields that are spherical while others are
deformed.  Consequently, one might expect to find intermediate situations in
which the shape undergoes rather large fluctuations away from the equilibrium
shape.  
In addition to the modes which have classical analogs, in the nuclear spectra
vibrational modes unique in a quantal system, such as those involving charge
exchange or excitation of the nucleonic spins or oscillations in the pair field,
are known and studied.

In the following of this section we take shape oscillations as an example of
particle-vibration coupling,
because experimental study of various properties of them has been carried out
for years and the resulting data are accumulated.
The density variations associated 
with the vibrational
motion make corresponding variations in the average potential.  The distortion
of the average potential gives a coupling between the degrees of
freedom of the vibration and individual particles \cite{BM75}, 
\begin{equation}
\delta V = - k_{\lambda}(r) \, \sum_{\mu} \, Y^{*}_{\lambda \mu}(\theta, \phi)
\, \alpha_{\lambda \mu}
\label{eq:pvc}
\end{equation}
where 
\begin{equation}
k_{\lambda}(r) = R_{0} \, \frac{\partial V(r)}{\partial r} \qquad \mbox{or} 
\qquad r \, \frac{\partial V(r)}{\partial r}
\label{eq:pvc-kr}
\end{equation}
where $R_{0}$ expresses the radius of the average static potential, $V(r)$, 
for which we use  
the form of Woods-Saxon potential. Either radial form on the r.h.s. of 
(\ref{eq:pvc-kr}) that is used in various publications  
gives nearly the same numerical results because the quantity 
$\partial V(r) / \partial r$  is concentrated on the nuclear surface, 
though the first form can be more reasonable.

One may consider a number of effects which arise from the coupling:  
For example,
the renormalization of the properties of both particles and vibrations,
or a self-consistent description of the vibrational motion itself 
in terms of 
one-particle degree of freedom, or the effect of the exclusion
principle between the degrees of freedom of particles and those involved in the
vibrational modes, as well as the orthogonality of different modes.  
In particular, in the case of the coupling being weak enough to be treated by
perturbation, one can systematically calculate those various effects, and 
the comparison between the calculated quantities and the observed ones provides
a quantitative information on the validity of our understanding of 
the structure of the basic particle-vibration coupling.    

Particle-vibration coupling in low-lying quadrupole vibrations is
often too strong to be quantitatively treated by perturbation.   
In contrast, low-lying octupole vibrations provide the data, by which our 
understanding can be quantitatively tested.  A beautiful example
is related to 
the octupole vibration of the doubly magic nucleus, $^{208}_{82}$Pb$_{126}$.  
There are many kinds of data on octupole vibrations in neighboring nuclei 
of $^{208}$Pb \cite{IH74},
which were nicely interpreted in terms of the particle-vibration coupling. 
In the following we take the septuplet of the octupole vibration, 
$[(h_{9/2} \, 3^{-})_{I}, I^{\pi}=3/2^{+}, ..., 15/2^{+}]$, observed in
$^{209}_{83}$Bi$_{126}$, as a beautiful example of the 
particle-vibration coupling.   
In Fig. 1 observed low-lying energy spectra of $^{208}_{82}$Pb$_{126}$ and
$^{209}_{83}$Bi$_{126}$ that are relevant to the present discussion are shown. 

\begin{figure}
\includegraphics[width=1.3\columnwidth]{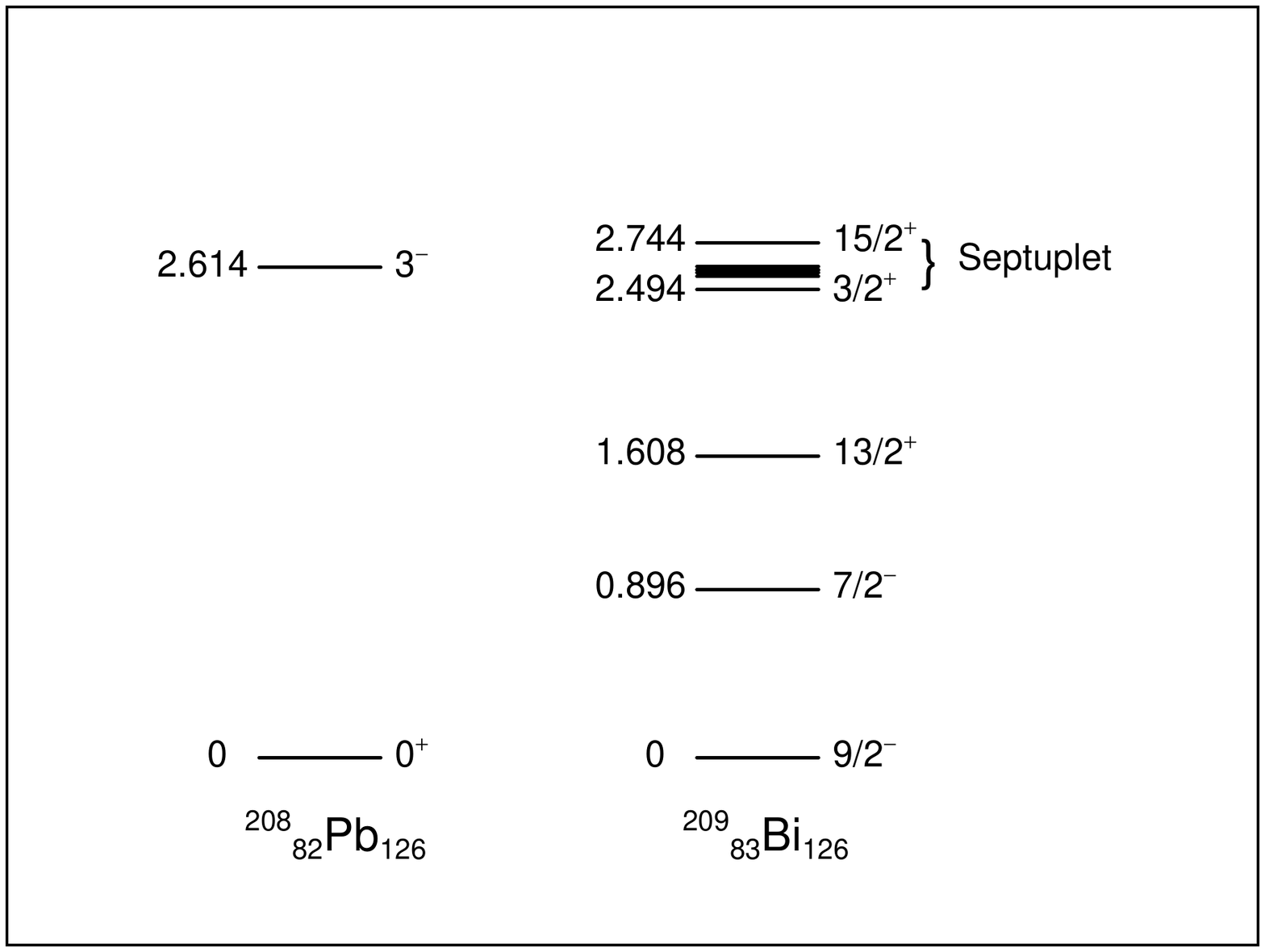}
\caption{
Observed low-lying energy spectra of $^{208}_{82}$Pb$_{126}$ and
$^{209}_{83}$Bi$_{126}$ that are relevant to the present discussion.
Excitation energies are expressed in MeV.}
\end{figure}

In the calculations of respective observed quantities shown in the following 
the contributions only  
in the lowest-order perturbation were
taken into account. However, it is important to note that 
all contributions in the lowest order are included when  
differential equations appropriate for respective contributions were 
numerically integrated instead of expanding the wave-functions 
in terms of a given finite basis.  

A number of beautiful experiments
on many faces of the septuplet members have been carried out especially 
during the 
sixties and the seventies, 
such as energies, decay scheme and octupole strength 
of respective members, 
one-nucleon transfer reaction cross section to populate some members.  
The properties of the octupole vibration itself
in $^{208}$Pb such as the transition density, quadrupole moment, and
double-phonon states were also explored.
Observed data in connection with this octupole vibration could be in almost all
cases treated by perturbation and gave a firm support for our
understanding of the shape (surface) oscillation and the related
particle-vibration coupling.     
The detailed comparison between experimental and calculated results can be 
found in available publications.  See, for example, \cite{IH74}. 
Therefore, in the following I show only two examples;
(i) energy shifts and
the decay scheme of the septuplet members of $^{209}$Bi and (ii) the radial
transition density of the octupole vibration of $^{208}$Pb.  

Parameters used as an input in the numerical calculations are the two observed 
quantities in $^{208}$Pb; the observed vibrational energy, 2.614 MeV, and 
the observed vibrational strength, $B(E3)$ = 32 $B_{W}(E3)$, which is used to
obtain the matrix element of $<n_{3}=1|\alpha_{3}|n_{3}=0>$.       
The value of 32 $B_{W}(E3)$ was taken from Coulomb excitation experiments
\cite{BP69, HKW72, GDH71}. 

In Table 1 calculated energy shifts (in the 2nd-order
perturbation of the particle-vibration coupling) and the decay scheme 
(in the 1st-order perturbation of the particle-vibration 
coupling) 
of the
septuplet members in $^{209}$Bi  are compared with experimental ones.   
The observed energy splitting of the
septuplet is 250 keV compared with 350 keV of the calculated one.  We note 
a remarkable
agreement between the calculated and experimental decay schemes, in
particular, six $B(E1)$-values obtained 
using $[e^{p}_{eff}(E1)]^{2} = 0.14 \, e^{2}$, 
while it is known to be 
extremely difficult to predict $B(E1)$ values 
in strongly-hindered low-energy E1 transitions in nuclei.   
This agreement may be used for supporting our basic understanding of the present
particle-vibration coupling and the surface vibration.  In addition, the obtained
value of $e^{p}_{eff}(E1)$ agrees approximately 
with the value \cite{BM75} estimated by taking into account the
reduction coming from both the center of mass motion and the presence of 
the isovector giant dipole resonance.  

\begin{figure} [ht]
\includegraphics[width=\columnwidth]{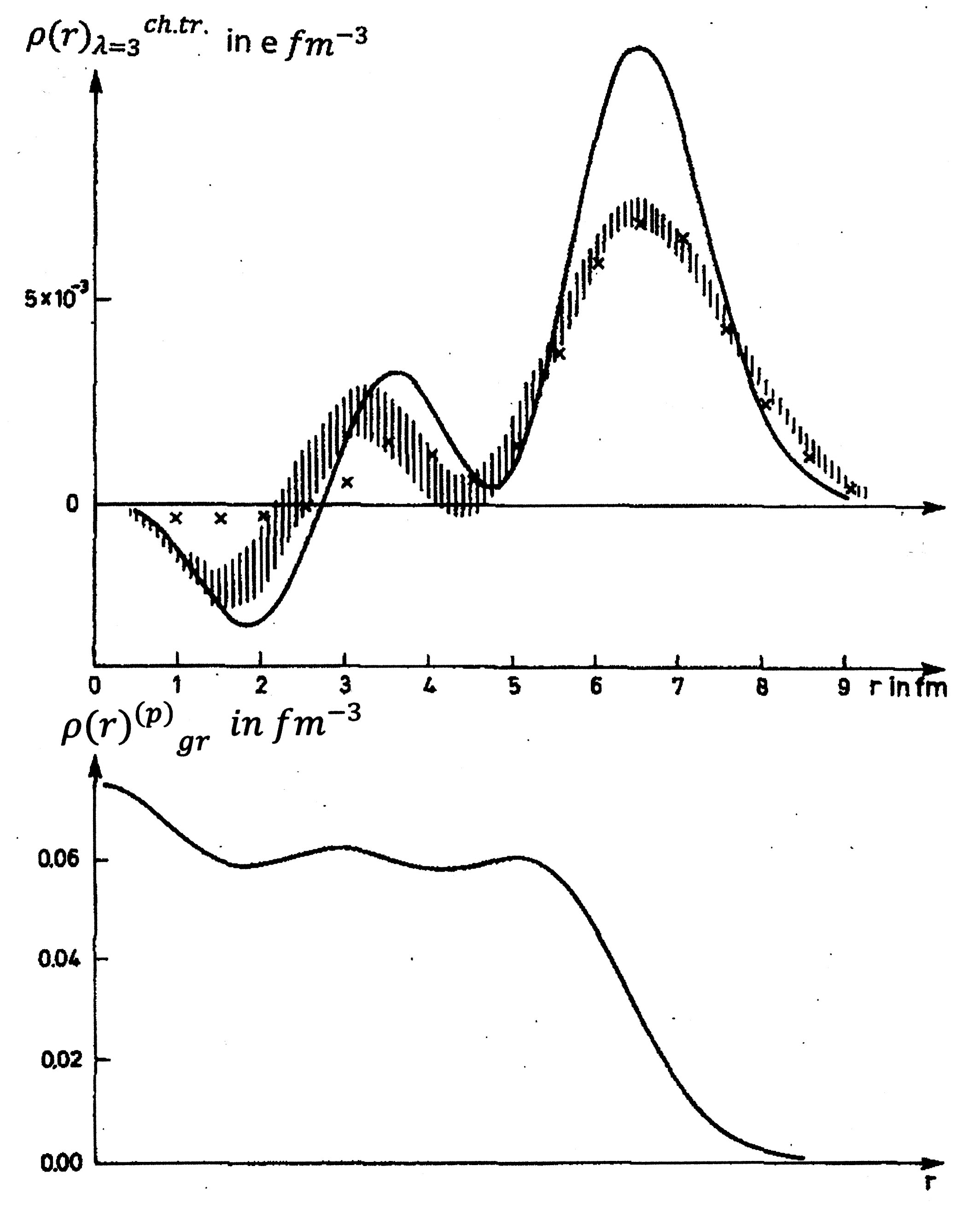}
\caption{\label{fig.2}
The upper part expresses the radial transition charge density of 
the 3$^{-}$ state at 2.614 MeV 
in $^{208}$Pb.  The solid line shows the calculated value \cite{IH77}, 
while for the
experimental values the dashed band was taken from Ref. \cite{MN72} and the
crosspoints were from Ref. \cite{JH70}. 
The lower part of the figure denotes the calculated charge density of the ground
state of $^{208}$Pb.  The scale of the x-axis in the lower part is the same as
the one in the upper part.  The figure is adapted from Fig. 1 of 
Ref. \cite{IH77}. 
}
\end{figure}

In Fig. 2 the calculated radial transition density 
(in the 1st-order perturbation of
the particle-vibration coupling) of the 3$^{-}$ state of $^{208}$Pb \cite{IH77} 
is compared with those obtained from electron inelastic scatterings.  
The large peak around the surface can be understood as a feature of the
shape oscillation.  The position of the first minimum around 5 fm coincides
closely with the $r$-value where the ground-state density reaches the maximum, 
noting that 
the radial transition density is approximately proportional to the radial
derivative of the static density.  The structure of the two small peaks inside
the nucleus depends on the microscopic shell-structure of the octupole vibration
in
addition to the general quantum-mechanical oscillatory structure of the static
density.

Another example in somewhat lighter-mass region, which exhibits a very similar
coupling between particles and an octupole vibration, was reported in 1982
\cite{MP82}.  
The septuplet members in $^{147}_{64}$Gd$_{83}$, 
$(f_{7/2} \, 3^{-})_{I^{+}}$ with  
$I$ = 1/2, ..., 13/2, in which the octupole vibration
3$^{-}$ of $^{146}$Gd at $E_{ex}$ = 1.579 MeV with $B(E3)$=$37 B_{W}(E3)$ was  
coupled to the 83rd odd-neutron in the $f_{7/2}$ orbit, 
were identified except the $I^{\pi} = 5/2^{+}$ member. 
Due to the dominance of the proton  
$(h_{11/2} \, d_{5/2}^{-1})_{3^{-}}$ component in the 3$^{-}$ state, 
not all properties of the septuplet members could be analyzed by
perturbation.  However, it is unique in the case of the octupole vibration 
of $^{146}$Gd that the two-octupole-phonon angular-momentum-stretched state, 
[($\nu f_{7/2}) \, (3^{-})^{2}$]$_{19/2^{+}}$, 
in $^{147}$Gd was experimentally identified and the
pretty strong anharmonicity in both the energy and the E3 transition rate was
subsequently analyzed \cite{PK82}.

\section{One-particle motion in deformed and rotating potentials, and 
yrast spectroscopy in high-spin physics} 
The very basic description of nuclear many-body systems is the
(self-consistent) mean-field approximation to the many-body problem.  In
particular, shape is the property of mean field.  
The ground states of some nuclei are described by densities and mean fields that
are spherical, while others are deformed.  In Fig. 3 the even-even
nuclei, of which the ground state is observed to be deformed, are shown.  
The deformation results from
one-particle shell-structure, namely bunching of one-particle levels, 
in contrast to the prediction of the liquid-drop
model.  For the nucleus as a quantum many-body system the presence of
deformation is a necessary condition for the occurrence of collective rotation,
and the symmetry of the deformation determines the structure of rotation.  The
most important deformation around the ground states of nuclei is the
axially-symmetric quadrupole deformation corresponding to a spheroidal shape
(cigar or pancake).  Axially-symmetric nuclei cannot perform a collective
rotation about the symmetry axis.

If some nuclei show deformation with a given symmetry it is simplest and most
convenient to start with the mean field, which has the same symmetry.  
One-particle motion in quadrupole-deformed potentials was solved in the fifties
almost at the same time by three groups \cite{SGN55,SAM55,KG56}.  
Among them, ''Nilsson model'' \cite{SGN55} 
has been used by many people in the analysis of data, 
presumably because the application of the Nilsson model using modified
oscillator potentials to actual
nuclei was easiest and very practical.    
The spectrum of one-particle orbits as a function of deformation, so-called 
''Nilsson
diagram'', has played an invaluable role in the study of deformed nuclei.  
(In the present article the word, ''Nilsson diagram'', 
is used for the diagrams of
one-particle energies as a function of quadrupole deformation using general
one-body potentials such as Woods-Saxon or HF potentials, and not only for the
diagram drawn by using modified oscillator potentials.)
Nilsson diagrams have been extremely useful for analyzing the data 
on deformed nuclei.    
This is mainly because the major part of the ($Y_{20}^{\ast} \cdot Y_{20}$) 
channel of the two-body 
quadrupole-quadrupole interaction is absorbed into 
the mean field.  
Consequently, the picture of one-particle 
motion in the deformed potential works for deformed nuclei 
much better than that of one-particle motion in spherical potentials 
for spherical nuclei.    
For example, see the successful quantitative analysis of various kinds of
experimental data of well-deformed (especially odd-A) nuclei such as 
$^{25}$Mg, $^{25}$Al, $^{159}$Tb, $^{169}$Tm, $^{175}$Yb, $^{175}$Lu, 
$^{177}$Lu, $^{177}$Hf, $^{235}$U, $^{237}$Np, $^{239}$Pu presented in Ref.
\cite{BM75}.      

The field of nuclear high-spin physics was activated 
by the observation of the ''back-bending'' phenomena \cite {AJ71}, 
which was the first
band-crossing (between the ground-band and the S-band) 
along the yrast line in well-deformed collectively-rotating
nuclei.  
During the seventies the possibility of studying nuclei with very large
angular-momentum was opened up, thanks to the facilities of accelerating 
heavy-ions.  Collisions between two heavy nuclei could 
produce meta-stable compound
systems with large angular-momenta.   Indeed, in heavy-ion collisions in the
late seventies 
it was possible to produce compound nuclei with angular momentum all
the way up to the limit set by the fission instability.  
The maximum value of angular momentum which nuclei can accommodate 
is the order of 100$\, \hbar$ for nuclei with $A \approx 130$
\cite{CPS74}.
The study of rapidly rotating nuclei provided the opportunity for exploring new
aspects of nuclear dynamics \cite{AB75,BM79,IH85,GHH86}.     

When the angular momentum is accommodated by a quantal system such as a nucleus,
one recognizes that there are two fundamental ways of building the angular
momentum \cite{BM74}.  The first one is collective rotation in the presence of
a significant amount of deformation, while the second one is to build the
angular momentum by spin alignments of individual particles when the angular
momentum is directed along an axis of symmetry of the nucleus.  In actual nuclei
we encounter situations in which the increment of angular momentum is achieved
partly by increasing the frequency of collective rotation and partly by
rearranging the occupation of one-particle orbits (namely, by increasing the
particle alignments).  In fact, the collective rotational spectra together with
the crossing of the ground band with the S-band are characteristic of this kind
of situation.  

Special interest in nuclear high-spin states attaches to the region in the
neighborhood of the yrast line representing the lowest energy for a given
angular-momentum.  In this yrast region the nucleus is cold in the sense that
almost the entire excitation energy of the nucleus is consumed in generating the
angular momentum.  Therefore, the structure in the yrast region is ordered with
simple excitation modes, and the study of those may be expected to give
important nuclear-structure information on how the nucleus responds to the large
centrifugal forces associated with rotation.  The path that the yrast line of
actual nuclei will follow in deformation space with increasing angular-momentum
will result from the competition between the macroscopic centrifugal distortion
effect and the quantal effects associated with shell-structure.  

In the yrast spectroscopy of medium-heavy nuclei 
an important role has been played by particles in
high-j orbits, such as 1g$_{9/2}$-, 1h$_{11/2}$-, 1i$_{13/2}$- and  
1j$_{15/2}$-orbits.  This is because the parity of high-j orbits is 
different from that of 
other one-particle orbits in respective major shells, therefore,
both in quadrupole deformation and under rotation particles in high-j orbits 
do not mix with those in  
neighboring orbits.  That means, the wave functions of particles in high-j 
orbits
have less ambiguity.  Furthermore, as soon as rotation sets in, particles in
high-j orbits start to align due to the particularly large Coriolis coupling,
while particles in neighboring other orbits still contribute to 
collective rotation.   
Consequently, the states consisting of high-j-shell configurations easily 
appear in the neighborhood of the yrast line.   
A typical example is that the intrinsic structure of the S-band contains 
a large
alignment coming from the aligned two quasiparticles in the relevant high-j
orbit, in contrast to the absence of such alignment in the ground band, at
the angular momentum of the band crossing.  

The basis for the analysis of shell-structure effects 
on nuclear high-spin states
is the study of one-particle motion in rotating potentials.  
The importance of understanding the physics in the yrast region, especially in
terms of one-particle motion in rotating potentials, was 
repeatedly emphasized by Bohr and Mottelson during the seventies.  
The cranking model
is easily treated and the rotational frequency appears as an explicit parameter.
Indeed, the major part of the analysis of nuclear high-spin phenomena has been
carried out in terms of cranking models of various types.  
In particular, the so-called Routhian diagram, namely the diagram plotting one
quasiparticle energies in the rotating frame 
as a function of cranking frequency, played for years a
central role in the analysis of nuclear high-spin data.  
I remember that around the middle of the seventies Aage Bohr was repeating 
to say to
us working at the Niels Bohr Institute that one should draw and study the
Routhian diagram for fixed deformation and pairing parameters.  
(At that time I stupidly thought that such simplified Routhian diagrams 
might be 
qualitatively excellent for getting ideas and ways of understanding physics,
but they would hardly be useful for practical numerical applications.)
In this connection, I later realized that, for example, 
without plotting such Routhian diagrams it would have been indeed
difficult to notice the fact that the interaction strength between the ground
band and the S-band, namely the sharpness of backbending phenomena, is an
oscillating function of the degree of high-j-shell filling \cite{BHM78}.  
The extended study with numerical works using realistic configuration spaces was
eventually carried out by R. Bengtsson and S. Frauendorf \cite{BF79}, and 
the Routhian diagrams of such a simple type turned out to be extremely useful
and have been successfully used in the analysis of
high-spin data for the following years.

However, one should keep in mind that the uniform rotation of a system, 
which is
a basic assumption of the cranking model, may become a poor approximation under
certain circumstances, because the presence of fluctuations in the collective
rotational frequency is recognized if one treats more reasonably the exchange of
angular momentum between individual particles and the potential produced by the
rest of the system \cite{IH85BM}.      
An example outside the applicability of the cranking
model in its simplest form is the description of the crossings 
of the bands with a
large difference of spin alignments, irrespective of whether the parameters in
the model are chosen self-consistently or not.  The inapplicability comes from
the fact that a mixing of the two bands for a given rotational frequency is 
basically considered in the cranking model, while the two bands should interact for a
given angular momentum \cite{IH76}.

The experimental front of high-spin physics made a tremendous progress in the
decades after the observation of ''back-bending'', especially due to the
development of multi-gamma-ray detectors, which had 
a resolution of orders greater
than that of the last generation, and $4\pi$ or $8\pi$ spectrometers; NORDBALL
and TESSA $\rightarrow$ EUROGAM and GASP $\rightarrow$ EUROBALL and GAMMASPHERE.

New frontiers were opened in the nuclear high-spin physics, when the
superdeformed band in $^{152}_{66}$Dy$_{86}$ was reported \cite{PT86} 
by finally observing gamma-ray spectra of discrete transitions in the
decay sequence of the superdeformed band of $^{152}$Dy 
extending to $I \approx 60$ $\hbar$,
thanks to the development of detectors.  
The observation of the discrete gamma-rays could directly and clearly show the
presence of the superdeformed band, compared with the information extracted
previously from the painstaking analysis of complicated continuum gamma-rays. 
It is noted that the largest angular
momentum, which was reported in discrete line studies before the discovery of
the superdeformed band, was only 46 $\hbar$.  Because of the large moment of
inertia of superdeformed nuclei, the large angular-momentum needed for the study
of such systems does not necessarily mean large rotational frequencies.  
Consequently, the unique behavior of particles in high-j orbits could largely
remain also in the analysis of superdeformed bands.  
The shape of superdeformation in nuclei, where the axis ratio of the
prolately-deformed shape is approximately 2:1, was known already in the sixties 
as (low-spin) fission isomers in actinide nuclei \cite{SMP62} 
and was interpreted 
as the result of the  
shell-structure which appears first for much larger quadrupole-deformation 
($\beta \approx 0.6$ in medium-heavy nuclei) than 
the one known in  nuclear ground states.  
In subsequent years the observation of super- or highly-deformed bands 
was reported also in
other mass region than $A$=150, such as $A$=130 \cite{NT88} and 190 \cite{JK91},
though the relevant shape was not always close to 2:1.

Before 1975 the view might have been accepted that there were nuclei that
were statically deformed and they were located far away from closed shells, 
while there
were nuclei that were spherical and they were located in the neighborhood of 
closed-shell regions, though in some nuclei located on the border of spherical 
and deformed nuclei the simultaneous presence of the low-lying states 
with spherical and deformed shapes  
was already recognized.  Thanks to the marvelous development of experimental
techniques in high-spin physics, collective rotational bands were found
at the excitation energy of several MeV 
also in some nuclei such as $_{82}$Pb, $_{50}$Sn and $_{20}$Ca isotopes
\cite{HW11},  
which are at least semi-magic and were supposed to be
spherical.  While it has been known that for a given nucleus Hartree-Fock
calculations often predicted local energy-minima at several different 
deformations, 
the possible presence of such several shapes in a given nucleus has
been indeed confirmed thanks to the modern experimental technique.    

In the present special issue of Physica Scripta I expect a number of
contributions by specialists in the field of high-spin physics.   Therefore, 
to those contributions I would leave to show 
a tremendous amount of beautiful results
obtained from all detailed numerical works in comparison with exciting and
impressive experimental data.

\section{Triaxial shape in nuclei}  
Theoretically the existence of nuclear deformation other than axially-symmetric
deformation was predicted, and it has been a great challenge to find
the predicted axially-asymmetric (triaxial) deformation of nuclei.  
In order to pin down triaxial deformation, it is essential to find 
the phenomena which are unique in axially-asymmetric shape.  In connection with
available experimental data on ''high-j'' configurations in the yrast
spectroscopy, some phenomena such as the signature-dependence of 
B(E2:I $\rightarrow$ I$-$1) values in odd-A nuclei 
and anomalous signature-splitting of
Routhians in odd-A and odd-odd nuclei, were theoretically suggested as the 
evidence for triaxial shape \cite{IH90}.  However, 
before 2000 a very clear and firm evidence
for stable triaxial shape was hardly obtained experimentally.  

On the other hand, the two
phenomena, wobbling excitation modes and chiral bands, 
are unique in triaxial shape and were intensively searched for 
by using the advanced technique developed in high-spin
physics.   Observation of beautiful wobbling
excitation modes was reported in the beginning of the 21st century, while, in my
opinion, experimental data which clearly pin down chiral bands 
have not yet been
obtained though several candidates for chiral bands have been reported.  
In this section I describe the present understanding of these
two phenomena.  

\subsection{Quantized wobbling observed in nuclei}  
Nuclei with a triaxial shape can rotate about any of the principal axes showing
rich spectra of collective rotation.  Though the rotation about the axis with
the largest moment of inertia is energetically cheapest, while freezing the
intrinsic structure a series of rotational bands can be built by transferring
some angular momentum to the other two axes.  The family of the rotational bands
is formulated in terms of vibrational excitations.  The classical analog of this
wobbling motion is the spinning motion of an asymmetric top, but the motion in
the nuclear system is quantized and expressed in terms of the wobbling phonon
number ($n_{W}$).  A family of rotational bands with wobbling excitations 
can be pinned
down by specific electromagnetic decay properties between them.  This quantized
wobbling phonon picture was first proposed by A. Bohr and B. R. Mottelson
described in \cite{BM75}.  

In 2001 such a wobbling band, a one-phonon wobbling excitation, was discovered
in the nucleus $^{163}_{71}$Lu$_{92}$ \cite{SWO01}.  One year later, in 2002 the
next wobbling excitation, a two-phonon rotational band, was reported in the same
nucleus \cite{DRJ02}.  In this experiment high spin states of $^{163}$Lu were
populated using a $^{29}$Si beam and a $^{139}$La target.  The emitted
gamma-rays were measured with the EUROBALL detector system.  The electromagnetic
nature of the decay transitions from the newly observed band is analyzed and the
results clearly show the two-phonon nature of the wobbling motion in good
agreement with calculations, in which the intrinsic degree of freedom of the
rotational bands in $^{163}$Lu is represented by one highly-aligned quasiproton
in the $i_{13/2}$ shell \cite{IH02}.   

The state with high-j aligned particles favors a specific (triaxial) shape
\cite{HM83} depending on the degree of the j-shell filling.   The 
$\gamma$-value favored by the aligned i$_{13/2}$ proton in $^{163}$Lu is 
around +20$^{\circ}$ (in the so-called ''Lund convention'' \cite{GA76}).   
States with
large alignments can easily appear in the neighborhood of the yrast line because
of the relatively small rotational angular-momentum (and thereby small
rotational energy) needed for building a given total angular-momentum.  
Furthermore, using a particle-rotor model, in which one quasiparticle of
i$_{13/2}$ protons is coupled to a triaxial core, it was shown \cite{IH87} 
that in the presence of one
high-j aligned quasiparticle and for the $\gamma$-value strongly favored by the
fully-aligned high-j particle the wobbling excitation of the collective
rotational angular-momentum of the core appears as the yrare band in a certain
range of angular momentum  
and the electromagnetic transitions between the yrast and yrare bands 
show a unique
pattern.

The two-phonon wobbling excitation observed in $^{163}$Lu is one of the most
exotic properties of the spinning nucleus.  The observation has been 
possible not least due to
an increased efficiency in the detection of gamma-rays with high resolution
germanium spectrometers which have been developed and became available during
the last
decades of the 20th century.  The observation of a two-phonon excitation, of
which the relevant degree of freedom is collective rotation, amplifies the
uniqueness of the finding for nuclear wobbling.  The nucleus has revealed its
exploitation of the quantal wobbling degree of freedom which proves the
existence of triaxial nuclei and adds a new dimension to the description of a
rotating nucleus.  

Since 2002 the observation of the rotational bands, 
which showed spectroscopic properties (though 
often only 
energies) very similar to those of $^{163}$Lu, were reported in neighboring
nuclei including $^{161, 165, 167}$Lu, indicating the presence of the wobbling
bands also in those nuclei.  
However, the original data on the nucleus $^{163}$Lu are so far most
beautiful and best to pin down the characteristic features of not
only the energies and spin-parity but also electromagnetic properties and the
two-phonon wobbling excitation.  Therefore, in the following I briefly describe
the observed data on $^{163}$Lu and their interpretation.   

The essence of the experimental data on the wobbling excitations of $^{163}$Lu
\cite{SWO01, DRJ02, GBH04} 
is shown in Figs. 4 and 5, compared with the calculated results obtained by
using the particle-rotor model, in which one high-j quasiparticle is coupled to
the core of triaxial shape \cite{SWO01, IH02, DHH02}.  
The moments of
inertia and alignments of the three TSD (Triaxial, Strongly Deformed) bands,
TSD1, TSD2 and TSD3, are
nearly identical, and the intrinsic structure of those TSD bands is understood
as containing an aligned high-j (= i$_{13/2}$) proton.  TSD2 is identified as
the one-phonon ($n_{W} = 1$) wobbling band built 
on the yrast TSD1 ($n_{W} = 0$), while TSD3 as the 
two-phonon ($n_{W} = 2$) 
wobbling band.  The identification of the two-phonon band is based on, among
others, the unusually large (at this high-spin) 
B(E2; TSD3 $\rightarrow$ TSD2) value in agreement with both 
the phonon picture and the calculated result of the particle-rotor model
\cite{IH02}.  
It is also noted that very small values of 
B(E2; TSD3,$I$ $\rightarrow$ TSD1,$I-2$) are in agreement with the assignment of
TSD3 and TDS1 as $n_{W}$=2 and 0, respectively.  
The $\gamma$-value in the calculation 
($\gamma \approx$ +20$^{\circ}$) was fixed by the observed ratios, 
B(E2; TSD2,I $\rightarrow$ TSD1,I$-$1)/B(E2; TSD2,I $\rightarrow$ TSD2, I$-$2) 
and 
B(E2; TSD3,I $\rightarrow$ TSD2, I$-$1)/B(E2; TSD3,I $\rightarrow$ TSD3,I$-$2), 
which are insensitive to used values of moments of inertia but are strongly
increasing functions of $\gamma$, especially for $\gamma \geq +20^{\circ}$.  
It should be also mentioned that 
for nuclei with Z$\approx$71 and N$\approx$94 ''ultimate
cranker' calculations \cite{HSP95} predicted triaxial shapes ($\gamma \approx
\pm 20^{\circ}$) with large quadrupole deformations ($\varepsilon_{2} \approx
0.38$) for all combinations of parity and signature in the region of angular
momenta, which are relevant for the observed TSD bands of $^{163}$Lu.  
The local minimum with $\gamma > 0$ is generally lowest, and at the minimum the
i$_{13/2}$ orbital is lowest in energy of the proton system with the favored
signature $\alpha_{f} = +$1/2 where $I$ = $\alpha \; mod \; 2$.

M1 transitions between $\Delta n_{W} = 1$ bands are in general strongly reduced
as seen in Fig. 5.  This is because in the text-book example \cite{BM75}  
namely in the absence of the high-j aligned particle, 
the total angular-momentum $\vec{I}$ is
the only vector in the system, thus, the magnetic dipole operator is
proportional to $g \vec{I}$.  
Then, M1 transitions should vanish 
in the case of the isotropic $g$-factor in the
body-fixed system.  
In the presence of fully-aligned 
intrinsic angular-momentum $\vec{j}$ and when possible quantum-fluctuations are
neglected,
the argument goes in a similar way to the above case when  
$\vec{I}$ above is replaced by $\vec{R}$ where $\vec{R}$ expresses 
the rotational angular-momentum of the even-even core. 
In contrast, the B(E2; $ n_{W}, I \rightarrow n_{W} \pm 1, I-1)$ value is
proportional to $I^{-1}$.  
The observed magnitudes as well as the zigzag pattern 
of both B(M1) and B(E2) values are in good agreement with the
wobbling picture obtained from the model consisting of one $i_{13/2}$
quasiparticle 
coupled to the triaxial-rotor. 
In particular,  
it is interesting to note that in Fig. 5 the zigzag pattern of 
both B(E2) and B(M1) values of 
$\Delta I = 1$ transitions in the wobbling regime is opposite to the one  
in the cranking regime.    
Furthermore, in the cranking regime one expects that for $\Delta I = 1$
transitions at high spins the B(M1) values are the order of unity, 
while the B(E2)values are the order of $I^{-2}$. Thus, it is seen that 
the expectation from the
cranking model totally disagrees with
the observation.

\subsection{Chiral bands}
Spontaneous formation of handedness or chirality is a subject of general
interest in molecular physics, the characterization of elementary particles, and
in optical physics.  The occurrence of chirality in a nuclear structure was
considered theoretically \cite{SF97} and, subsequently, experimental
level schemes in
some odd-odd nuclei (for example, $^{134}_{59}$Pr$_{75}$) 
exhibiting the patterns similar to the predicted ones 
have been reported \cite{KS01}.

The total Hamiltonian for the nuclear system is taken to be invariant under the
exchange of the right- and left-handed geometry.  Chirality in triaxial nuclei
is characterized by the presence of three angular-momentum vectors, which are
generally noncoplanar and thereby make it possible to define chirality.  One of
the three angular-momenta is the collective rotational angular-momentum, while
the other two in odd-odd nuclei are,
 in practice, angular-momenta of quasiproton and 
quasineutron in high-j orbits.  The hallmark of nuclear chirality is the
observation of two almost degenerate $\Delta I = 1$ rotational bands, chiral
bands, having the same parity.  Those almost degenerate states are expected to
appear only after an appreciable amount of collective rotation develops, while
for higher rotational frequency the basis for chirality, namely the noncoplanar 
structure of the
three angular-momenta, will be destroyed.  The reason is that 
in order to make a
given large total angular-momentum, it is cheapest to align all constituent
angular-momenta to the direction of collective rotation.
Thus, chiral bands may be found only in a limited intermediate 
region of angular-momentum.

Observed so-called chiral bands, two $\Delta I$ = 1  bands, are typically a few
hundreds keV apart and one must find a reasonable explanation of this energy
difference.  Furthermore, though the presence of two close-lying bands 
may indicate chiral geometry,
this geometry can be pinned down in a more definitive way if electromagnetic
transition probabilities expected for the chiral bands are experimentally
confirmed.  The trivial behavior of electromagnetic transition probabilities 
or the
''trivial selection-rule'' expected in general chiral pair-bands is: 
''For the states with $I \gg 1$ the corresponding probabilities and moments 
in the two bands should be identical or in practice almost the same.''    
Namely, when chiral geometry is realized, observed two chiral-degenerate states
may be written as 
\begin{eqnarray}
| I+ \rangle & = & \frac{1}{\sqrt{2}} \left( |IL \rangle + |IR \rangle
\right)  \\ 
| I- \rangle & = & \frac{i}{\sqrt{2}} \left( |IL \rangle - |IR \rangle
\right) 
\end{eqnarray}
where left- and right-handed geometry states are denoted by $|IL \rangle$ and 
$|IR \rangle$, respectively.  For $I \gg 1$ it is expected that 
\begin{equation}
\langle IL | EM | IR \rangle \approx 0
\end{equation}
where $EM$ expresses electromagnetic operators.  Then, within the chiral
pair-bands one expects 
\begin{eqnarray}
B(EM; I_{1}+ \rightarrow I_{2}+) & \approx & B(EM; I_{1}- \rightarrow I_{2}-) \\
B(EM; I_{1}+ \rightarrow I_{2}-) & \approx & B(EM; I_{1}- \rightarrow I_{2}+) 
\end{eqnarray}

In addition, 
in Ref. \cite{TK04} the selection rule derived from a special case of odd-odd
nuclei, in which the orbits of both odd-neutron and odd-proton are 
the same high-j,
was obtained by using the particle-rotor model in which  one-proton- and
one-neutron-quasiparticle were coupled to a triaxial core \cite{BM75}.  The
majority of observed two close-lying $\Delta I = 1$ bands in odd-odd nuclei 
around $N \approx 75$ 
\cite{KS01} belongs to this special case where j=h$_{11/2}$.  In this case, an
additional symmetry exists in the model Hamiltonian and 
an associated quantum number 
can be obtained.  Using this quantum number a further 
selection rule for electromagnetic transitions was derived 
in an unambiguous manner
\cite{TK04}.  To my knowledge, two close-lying $\Delta I = 1$ bands observed
in odd-odd nuclei, which
approximately satisfy this selection rule, are not yet found.    

As a matter of fact, we have not yet observed two close-lying $\Delta I$=1 
bands, of which electromagnetic transition probabilities approximately follow
even the ''trivial selection-rule''.  Thus, we have to say that 
in spite of the nice theoretical idea we have not yet obtained 
experimental data which show the presence of
chiral bands.  This fact may come from the transient character of chiral bands,
which can survive only in a certain range of angular momentum.  
The detailed description of our present knowledge about chiral bands together
with so far available experimental data will be presented in other contributions
to this special issue.

\section{Nuclear structure as neutron-drip-line approaches}  
An example of the systematic change of energies of one-particle orbits due to
weak binding 
which we learn in traditional textbooks is Thomas-Ehrman shift.   
The reduction of the Coulomb energy for the loosely-bound proton orbits 
and for the
unbound resonance states is called Thomas-Ehrman shift.
A typical example is: 
the rather large difference (370 keV) in excitation energies of 
the s$_{1/2}$ levels between 
mirror nuclei, E$_{x}$ = 0.87 MeV in $^{17}_{8}$O$_{9}$ 
where the neutron separation energy S(n)=4.14 MeV and E$_{x}$ = 0.50 MeV in 
$^{17}_{9}$F$_{8}$ where the proton separation energy S(p)=0.60 MeV, 
can be explained in terms of the reduction in Coulomb energy associated with the
loosely bound proton \cite{BM69}.  
In the case of protons the effect of weak binding on energies and wave functions
can be seen only in such very light nuclei because the height of Coulomb barrier
becomes increasingly high as Z increases. 
In the present section I confine my attention to the 
neutron-drip-line nuclei despite the significant interest that is also
associated with the proton-rich side (among others, nuclear astrophysical
importance, neutron-proton correlations in the Z$\approx$N region, new
opportunities provided by $p$ and $2p$ decays).  

The study of unstable nuclei, especially neutron-drip-line nuclei which contain 
very weakly-bound neutrons, has opened a new field 
in the research of the structure
of finite quantum-mechanical systems.  The study is important not only because
of the interests in nuclear astrophysics such as understanding the production of
energy and the synthesis of elements in stars and during stellar events, but
also because it provides the opportunity to learn the properties of fermion
systems with very loosely bound particles, some density of which can extend to
the region far outside the region of the main density of the system.  Because
the Fermi level of drip-line nuclei lies close to the continuum, both
weakly-bound and positive-energy one-particle levels play a crucial role in the
many-body correlations of those nuclei.  

Among various exciting phenomena which have been explored 
in the study of the structure of drip-line nuclei 
I pick up the following two topics in this section.     
First, we can find the systematic change of
neutron shell-structure, 
as S(n) decreases and approaches zero or even negative values (namely, 
one-particle resonances).  As a result of it, traditional magic numbers known 
for stable nuclei may be changed and, furthermore, 
nuclei with closed-shell configurations in the traditional stable nuclei 
may become 
deformed.  I describe our understanding of these phenomena 
in the following first subsection,
while in the second subsection I touch the  
halo phenomena in spherical and
deformed nuclei which can be found along   
the neutron-drip-line having some components of weakly-bound 
$\ell$=0 or 1 neutrons.

\subsection{Shell-structure change and deformation}
Recent experimental data obtained by using radioactive ion beams reveal that the
neutron numbers such as $N$=8, 20 and 28 are no longer magic numbers in some
nuclei toward the neutron-drip-line.  In traditional stable nuclei the
neutron separation energy $S(n)$ is typically 7-10 MeV.  Thus, the information on
one-particle shell-structure 
around the energy has been easily obtained experimentally. 
The prominent change of the level structure in the region of 
$\varepsilon_{j}(n) < 7$ MeV can be seen, for example, in Fig. 2-30 of
Ref. \cite{BM69}, where the energies of neutron orbits 
in spherical
Woods-Saxon potentials are shown.  
When the potential strength becomes weaker by decreasing the mass-number $A$,
thereby decreasing the radius of the potential, eigenvalues of all orbits 
$\varepsilon_{j}(n) < 0$ become less bound.  However, neutron-orbits with larger
$\ell$ (thus, larger $j$) lose the binding energy more rapidly than those with
smaller $(j \ell)$.  This is because due to the presence of higher centrifugal
barrier the major part of the wave functions of the orbits with larger 
$(j \ell)$ stays inside the potential and, thus, eigenvalues 
$\varepsilon_{j}(n)$ are more sensitive to the strength of the potential than
those of the orbits with smaller $(j \ell$).  
Taking a finite square-well potential as an example, the probability for bound
one-neutron wave-functions to remain inside the potential in the limit of
eigenvalues $\varepsilon_{n \ell} (<0) \rightarrow 0$ is tabulated in Table II.

Since the effective interactions, which can be reliably used 
in Hartree-Fock (HF) calculations of unstable nuclei far away from the stability
line, are not yet established, 
in the present article we use the Woods-Saxon potential for nuclear one-body
potential, of which parameters are taken from p.239 of Ref. \cite{BM69} unless
otherwise stated.

Weakening of the Woods-Saxon potential can be done by reducing either the
potential radius or the potential depth.  The two ways of weakening
change the shape of the potential in respective manners.  
In Fig. 6 an example of neutron one-particle
energies as a function of the depth of a spherical Woods-Saxon potential is
given.  It is noted that as one-particle energy $\varepsilon_{\ell j} (<0)$ 
approaches zero the $2s_{1/2}$ level approaches the $1d_{5/2}$ level and may
eventually become lower than the latter. Realizing this result the possible
neutron magic number N=16 was suggested in Ref. \cite{AO00}.  

In the following I use the simple argument: a large one-particle level 
density around the
Fermi level at the spherical point may lead to a deformation.  
The argument is based
on the following known fact: In very light nuclei the many-body pair-correlation
may be neglected in a good approximation. Then,
nuclei with a few nucleons outside a closed shell can be 
already deformed, because
using the near degeneracy of one-particle levels those nucleons have a
possibility of gaining energy by breaking spherical symmetry (Jahn-Teller
effect \cite{JT37}).

Due to the same physics 
mechanism as the nearly-degenerate $2s_{1/2}$ and $1d_{5/2}$ 
levels
in the case of weak binding as shown in Fig. 6, the $2p_{3/2}$ and $1f_{7/2}$ 
levels become nearly degenerate in the case of weak binding or low-lying
resonant levels \cite{IH07, IH12}.  
The Nilsson diagram based on a deformed Woods-Saxon potential 
which is appropriate
for $^{37}_{12}$Mg$_{25}$ \cite{IH07} is shown in Fig. 7.  The $1f_{5/2}$ and
$2p_{3/2}$ resonant levels for $\beta$=0 are found at +5.22 and +0.018 MeV with
the widths 2.08 and 0.005 MeV, respectively, where one-particle resonance in
deformed potentials is defined using the eigenphase formalism 
\cite{RGN66, IH05}.    
Using the eigenenergy 
$\varepsilon (1f_{7/2})$ = $-$0.66 MeV, the distance between the $1f_{7/2}$ and
$2p_{3/2}$ levels is 680 keV, which is very small compared with the distance
obtained in the case that both levels are well bound, as known 
from the presence
of the magic number N=28 in stable nuclei.  This near degeneracy of the 
$1f_{7/2}$ and $2p_{3/2}$ levels at $\beta$=0 directly means the
disappearance of the N=28 energy gap (or magic number) and leads to the fact 
that the 
N = 20-26  nuclei with weakly-bound neutrons in the $1f_{7/2}$-$2p_{3/2}$ shells
may prefer being deformed in the
case that the proton configuration allows the deformation \cite{IH07}.  

In recent experiments \cite{PD13} the even-even Mg-isotopes (N=22-26) towards
the neutron-drip-line are found to be indeed deformed by observing small values
of $E(2_{1}^{+})$ and possible 
$E(4_{1}^{+})/E(2_{1}^{+})$ ratios, while the 
odd-N nuclei, $^{31}_{10}$Ne$_{21}$ and 
$^{37}_{12}$Mg$_{25}$, are duly interpreted as deformed p-wave halo nuclei
\cite{TN09,NK14}.

\subsection{Spherical and deformed halo phenomena}  
Interests in nuclear halo phenomena were aroused 
by the observation of a remarkably large
interaction cross section of $^{11}_{3}$Li$_{8}$, which suggested a large
deformation and/or a long tail
in the matter distribution \cite{IT85}.  The long tail is later 
interpreted as a
two-neutron halo phenomenon.  
The observed neutron-halo structure makes it clear that the extreme
difference in the radial motion of weakly-bound $\ell$=0 and $\ell$=1 neutrons
from the radial distribution of the core particles indicates the
approximate decoupling of the halo particles from the core of the nuclear
system.  The effects of this decoupling on pairing, deformation, and collective
rotation is the interesting issue to be studied.  

The condition of the formation of neutron halo is 
that one or two least-bound neutrons
have small separation energies, say $S(n) < 1$ MeV, and some components of low
orbital angular-momentum, $\ell$=0 or 1.  
If it is one-neutron halo in spherical odd-N nuclei, the wave function of the
halo neutron has the $\ell$=0 or 1 component with the probability close to
unity.  If it is one-neutron halo in deformed odd-N nuclei, the probability of
$\ell$=0 or 1 component in the halo-neutron wave-function can be considerably
smaller than unity.  In the latter case the one-neutron wave-function can
contain considerable amounts of high-$\ell$ components, which are spatially
distributed within the same limited region as the well-bound even-even core
nucleus.  Thus, in the reactions to which extended tail exclusively contributes, 
for example Coulomb break-up reactions, one detects only the $\ell$=0 or
1 component of the one-neutron wave-function.  This seems to be the case of
Coulomb break-up reactions of $^{31}_{10}$Ne$_{21}$ \cite{TN09} and
$^{37}_{12}$Mg$_{25}$ \cite{NK14}.  Though the deformation of $^{31}$Ne
and $^{37}$Mg was predicted in 2007 \cite{IH07} as a result 
of the shell-structure of
neutron-drip-line nuclei described in the previous subsection, the
evidence for deformation of the core nuclei $^{30}$Ne and $^{36}$Mg was recently
reported.  See Refs. \cite{PF10} and \cite{PD13}, respectively.  

As is seen from the examples of $^{31}$Ne and $^{37}$Mg, if nuclei 
are deformed,  
one-neutron halo can be found in many more nuclei with different neutron-numbers
N than in the case that nuclei are limited to be spherical.  This is because
taking the example of $\ell$=0 halo, in the spherical case one-particle 
$s_{1/2}$-orbit is obtained only once in every $N_{ho}$=even major-shell, where
$N_{ho}$ expresses the principal quantum-number of the harmonic oscillator (ho).  
Therefore, spherical $\ell$=0 halo may be observed only at very special
drip-line nuclei, in which a neutron $s_{1/2}$-orbit lies 
around the Fermi level. 
In contrast,
in deformed nuclei all $\Omega^{\pi}$ = 1/2$^{+}$ intrinsic states acquire
$\ell$=0 components induced by the deformation and, thus, all those 
$\Omega^{\pi}$ = 1/2$^{+}$ orbits have 
a chance to make a deformed s-wave halo if the orbits are weakly bound \cite{HM03}. 
Needless to say, an
$\Omega^{\pi}$ = 1/2$^{+}$ orbit can be created from every positive-parity
orbits such as $s_{1/2}$, $d_{3/2}$, $d_{5/2}$, $g_{7/2}$, $g_{9/2}$,
.....   Furthermore, the $\ell$=0 component in a given $\Omega^{\pi}$ = 1/2$^{+}$
orbit increases as the eigenvalue $\varepsilon_{\Omega}$ ($<0$) approaches zero,
and the probability of the $\ell$=0 component approaches unity in the limit of
$\varepsilon_{\Omega} \rightarrow 0$ \cite{TM97, IH04}.  Similar comments apply
to the $\ell$=1 halo, because in the spherical case one-particle $p_{1/2}$- and
$p_{3/2}$-orbits are obtained only once in every $N_{ho}$=odd major shell.  In
contrast, in deformed nuclei $\Omega^{\pi}$ = 1/2$^{-}$ and 3/2$^{-}$
intrinsic states originating from every negative-parity orbits such as 
$p_{1/2}$, $p_{3/2}$, $f_{5/2}$, $f_{7/2}$,
$h_{9/2}$, $h_{11/2}$, ....... 
acquire $\ell$=1 components induced by the deformation, and all
these orbits may make deformed $\ell$=1 halo if the orbits are weakly bound.  
The $\ell$=1 component in given
$\Omega^{\pi}$ = 1/2$^{-}$ and 3/2$^{-}$ orbits 
increases also in the limit of $\varepsilon_{\Omega} (<0) \rightarrow 0$, 
but the
amount of increase depends on both one-particle orbits and potentials.

\section{Oblate ground state of even-even nuclei}
It is known that almost all known deformed even-even nuclei in the medium-heavy
mass region can be interpreted in terms of prolate axially-symmetric dominantly
quadrupole deformed shape.  
In the absence of pair correlation one obtains the number of prolate systems 
approximately equal to that of oblate ones in the simple models such as  
one-major-shell harmonic-oscillator or single-j shell.  However, 
when HF calculations with appropriate effective
interactions are performed in many well-bound nuclei, 
the dominance of prolate shape except
for very light nuclei is obtained in agreement with the experimental
observations.   In my opinion, the nature of the element
responsible for the overwhelming dominance of prolate shape has not yet been
adequately understood.    

Figure 6-48 of Ref. \cite{BM75} expresses the single-particle spectrum for 
axially-symmetric quadrupole-deformed oscillator potentials and 
the resulting magic numbers
for ($\omega_{\perp} : \omega_{3}$) = (1:2), (1:1) and (2:1).  The magic numbers
for spherical shape (1:1) are well known in textbooks for years.  On the other
hand, the prominent shell-structure with the prolate shape (2:1) has played an
important role in understanding the occurrence of both fission isomers
\cite{SMP62, BM75} and high-spin superdeformed bands.  These phenomena have 
provided striking evidence for the shell structure in nuclear potentials 
with much
larger deformations than those encountered in the ground states of heavy nuclei.
 Though neither spin-orbit potential nor surface effects, which are important
 elements in nuclear potentials, are present, the shell structure seen in the
 oscillator potential has helped us to understand the physics in a simple
 terminology.  In the present section it is shown that 
 the possible oblate shape of the ground states of 
 light well-bound nuclei, say $Z < 30$, can 
 also be easily understood in terms of the ''Nilsson diagram'' 
 based on the deformed
 oscillator potential.  

Recently, the neutron-rich nucleus $^{42}_{14}$Si$_{28}$ was reported to show a
rotational spectrum \cite{ST12} , namely the ratio of observed excitation
energies is $E(4_{1}^{+})/E(2_{1}^{+})$= 2.93(5) where 
$E(I_{1}^{\pi})$ expresses the
excitation energy of the lowest level with the spin-parity $I^{\pi}$, though the
spin-assignment of the $4^{+}$ state is not yet actually 
pinned down experimentally.  
The observed value of $E(2_{1}^{+})$, 770 keV, 
is not small compared with possibly prolate Mg isotopes ($^{34, 36, 38}$Mg), 
of which $E(2_{1}^{+})$ is around 650 keV.  The relatively low
moment of inertia corresponding to $E(2_{1}^{+}) =$ 770 keV  
may indicate an oblate deformation. 
One may wonder the reason why the nucleus with the neutron-number $N$=28 is
deformed and not spherical, as the neutron-number 28 is a well-known
magic-number in the j-j coupling shell-model and the observed 
neutron separation energy
of $^{42}$Si, $S(n)$= 3.6 MeV, is not small.

When even-even nuclei in the range of $6 \leq Z \leq 30$, of which the observed
electric quadrupole moment of the first-excited 2$^{+}$ state is clearly
positive corresponding to an oblate shape (or a fluctuation towards oblate
shape), are looked for, one finds only five nuclei: $^{12}_{6}$C$_{6}$, 
$^{28}_{14}$Si$_{14}$, 
$^{34}_{16}$S$_{18}$, $^{36}_{18}$S$_{18}$ and $^{64}_{28}$Ni$_{36}$.  
The proton and neutron numbers of these five nuclei remind us of the magic
numbers of oblate deformation with the frequency ratio 
($\omega_{\perp} : \omega_{3}$) = (1:2) and (2:3) in the deformed oscillator
potential. See Figure 6-48 of \cite{BM75}.  Namely, the magic numbers are 
Z = N = 6, 14, 26, 44, ... for the (1:2) deformation 
while N = Z = 6, 8, 14, 18, 
28, 34, 48, ... for the (2:3) deformation.
When the neutron and/or proton numbers are equal to one of those magic numbers,
the system in the deformed potential is supposed to be especially stable for
respective ($\omega_{\perp} : \omega_{3}$) deformations, though the total
deformation is determined by both proton and neutron numbers.  The deformation
parameter $\beta$ of the ground state obtained from experimental data is
relatively large for lighter nuclei, but certainly smaller than 
$\mid\beta\mid$ = 0.7.  The deformation parameter 
$\delta_{osc}$ = $(\omega_{\perp} - \omega_{3}) / \bar{\omega}$, where
$\bar{\omega} = (2\omega_{\perp} + \omega_{3}) / 3$, is equal to $-$0.43 and 
$-$0.75 for the ratio ($\omega_{\perp} : \omega_{3}$) = (1:2) and (2:3),
respectively.  Thus, except for extremely light nuclei 
the magic numbers for the 
(2:3) shape may be more realistic than those for the (1:2) shape, considering
$\delta_{osc} \approx \beta$.

In Fig. 8 the calculated Nilsson diagram for 
$^{42}_{14}$Si$_{28}$, which is obtained based on realistic
Woods-Saxon potentials, is shown \cite{IH14}.   
From Fig. 8 it is seen that the energy difference, 
$\varepsilon(2p_{3/2}) - \varepsilon(1f_{7/2})$, is only 1.65 MeV, which is 
smaller than the standard $N$=28 energy gap known in the j-j coupling shell model. 
Indeed, the energy gap on the oblate side, 
which is largest around $\beta = -0.4$, is appreciably larger than 
the gap at
$\beta = 0$.  Large energy gaps on the oblate side around 
$\beta = -0.4$ occur at $N$ = 14, 18 and 28.  Those neutron-numbers are in fact
exactly the magic numbers for the (2:3) deformation of the deformed oscillator
potential, though the Woods-Saxon potentials used for drawing Fig. 8 of course 
contain the
spin-orbit potential with the standard strength.  In contrast, 
it should be noted
that the neutron numbers, at which large energy gaps are found on the
prolate side of Fig. 8, such as $N$ = 12, 16, 24 and 28 around 
$\beta = +0.4$
have no relation with the magic numbers for the prolate 
($\omega_{\perp} : \omega_{3}$) = (3:2) deformation of the oscillator potential,
$N$ = ..., 14, 22, 26, 34, ...   See also Fig. 9 for the 
particle numbers in slightly heavier nuclei, 
at which large energy gaps are found in Nilsson diagrams
based on realistic potentials, in comparison with those based 
on the oscillator potential.

For $N$=28 on the oblate side of Fig. 8 four doubly-degenerate Nilsson levels
with $N_{ho}$=3 are occupied, while none
of $N_{ho}$=2 orbits are unoccupied.  
This is the same configuration as the one at the magic number 28 of the
(2:3) deformation of the oscillator potential.  
In short, the possible oblate deformation of $^{42}$Si can be understood as a
result of the combination of the facts: (i) Narrowing the spherical $N$=28 magic
number due to the shell-structure change in very neutron-rich nuclei; 
(ii) $N$=28 remains as ''a magic number'' for the moderate-size ($\beta \approx
-0.4$) oblate deformation in realistic nuclear potentials, and it is in fact  
the magic number for the oblate (2:3)
deformation of the oscillator potential; (iii) Oblate shape is much favored 
also by the proton number 
$Z$=14.  Note that the shell structure for protons in  
lighter well-bound nuclei is not so different from that 
for well-bound neutrons.  

The different correspondence between the realistic Woods-Saxon potential and the
oscillator potential for the prolate deformation 
from for the oblate deformation
seems to come mainly from the different behavior of the Nilsson one-particle
levels connected to the high-j shell (the $1f_{7/2}$ shell in the present case) 
on the prolate side from on the oblate side.  
The different behavior of the high-j
Nilsson levels was discussed in detail in \cite{HM09} in relation to the numbers
of oblate/prolate nuclei.  Due to the sign of nuclear spin-orbit potential
a given high-j orbit in a spherical potential 
is strongly pushed down relative to other orbits belonging to the same
oscillator major
shell.  Consequently, on the oblate side the unique shell-structure 
coming from the
presence of the high-j orbit in realistic potentials is disturbed 
soon after deformation sets in, and large energy gaps in the Nilsson diagram
occur at the particle numbers similar to those
in the oscillator potential. 
In contrast, on the prolate side the shell structure originating from the high-j
orbit in realistic potentials 
survives in the range of the realistic quadrupole deformation and,
thus, the particle number, at which a large energy gap occurs, is considerably
different from that of the oscillator potential.

Because of the simple property of the shell structure on the oblate side 
mentioned above, which is common to the Woods-Saxon potentials with realistic
parameters (and also realistic HF potentials), one may pretty reliably predict
the light nuclei, of which the ground state may have an oblate shape.  A good
candidate for the oblate shape of light nuclei, which is immediately
obtained from the above discussion, is $^{20}_{6}$C$_{14}$.

\section{Conclusions and discussions}
In my understanding the interplay between one-particle and collective degrees of
freedom has been a fundamental theme in the nuclear-structure physics of 
Bohr and
Mottelson.  In all topics described in the present article the interplay is
explicitly recognized.  Taking an example of particle-vibration coupling
presented in Sec. II, first
of all the coupling gives the relation between the vibration of an 
even-even nucleus
and that of the neighboring odd-A nuclei.  At the same time the coupling
specifies the structure of the vibration of the even-even nucleus 
itself such as the transition density, which in turn gives the octupole
vibrational strength.   
The self-consistent description of the vibration in terms of one-particle motion
has to be satisfied.  In the case that 
the coupling is weak enough to be treated by
perturbation, it is easy to quantitatively check the validity of 
our understanding of the
vibration.  

The unique role played by particles in
high-j orbits is noticeable in almost all topics.  Some unique features of
particles in high-j orbits are:
approximately high-j wave-functions not only for the practical size of 
quadrupole deformation but also
under an appreciable amount of 
rotation; large Coriolis coupling; strong alignment immediately after
rotation sets in; unique contribution to octupole vibrations; in case of full
alignment of one high-j quasiparticle a given $\gamma$ deformation is preferred 
as a function of shell-filling; unique 
contributions to the shell-structure for prolate shape and for oblate
shape, respectively; and so on.  

Though the idea of particle-vibration coupling itself can be applied to 
any kind
of vibrations in nuclei, numerical applications have been so far done
extensively in the isoscalar shape oscillations. 
When the vibration is the isoscalar quadrupole oscillation in spherical nuclei, 
the
associated particle-vibration coupling is often so strong that it cannot be
treated by perturbation.    
In the present article the particle-vibration coupling, in which 
the vibration is a typical 
isoscalar octupole shape-oscillation, the lowest excited 3$^{-}$  state of 
$^{208}_{82}$Pb$_{126}$, is chosen.  In this case most data can be analyzed by
the perturbation treatment of the coupling.  Thanks to the richness of 
available experimental data on the octupole-vibrations in neighboring nuclei of
$^{208}$Pb, I believe we have confirmed that Bohr and Mottelson's basic idea 
of the shape oscillations and the associated particle-vibration coupling 
in nuclei is definitely on the right track.  

Nuclear high-spin physics, in particular the yrast spectroscopy, was very
actively and successfully developed during the last three decades of the 20th
century, together with the tremendous progress in multi-gamma-ray detectors,
4$\pi$ or 8$\pi$ spectrometers and heavy-ion accelerators.  At the beginning
people were just fascinated by studying various kinds of rotational bands
obtained from the analysis of observed discrete $\gamma$-rays.  The analysis of
the data around the deformed ground-state was carried out based 
on Nilsson diagrams, while
at higher spins it was often done based on Routhian diagrams.  
A clear evidence for
Mottelson-Valatin effect \cite{MV60}, which was wondered once to be the origin 
of back-bending phenomena, was of course looked for.  However, a clear-cut
evidence for the phase transition might have been difficult to be seen 
in finite
systems such as nuclei.  After the discovery of the superdeformed (2:1) band   
in 1986, hyperdeformed (3:1) bands were looked for, but the analysis of
further complicated discrete and continuum 
$\gamma$-rays has been so far unsuccessful in
drawing any 
clear conclusion.  

Experimental findings of quantized wobbling modes are the
result of the technology developed in high-spin physics.   
The wobbling mode in $^{163}$Lu is the first one observed in nuclei, and it is
even now the most beautiful and complete one.  The presence of one aligned 
$i_{13/2}$ proton in the intrinsic configuration of $^{163}$Lu 
certainly helped the mode to
appear around the yrast line so as to be found more easily.  

The electromagnetic properties of observed two close-lying $\Delta I = 1$ 
rotational bands are being explored presently by several experimental groups. 
So far, whenever the electromagnetic transition probabilities were measured,
they did not follow even the ''trivial selection-rule''.  
However, if the close-lying $\Delta I = 1$ bands 
are not interpreted as chiral pair-bands, one may wonder what they are.  

Drip-line physics with halo phenomena is relatively newly-developed area in the
study of nuclear structure, as a result of radioactive ion beam facilities
constructed all over the world.  It has given us the opportunity to learn the
properties of fermion systems with very loosely bound particles.  The heaviest
odd-N halo nucleus so far known is $^{37}$Mg, which is interpreted as the 
deformed
p-wave halo.  In heavier nuclei the ratio of the fraction of the particles, 
which lie in the
extended region far outside the region of the main density of the system, 
to the total number of particles becomes smaller.  Nevertheless,  
the reactions such as Coulomb break-up, which are extremely sensitive to the
density far outside the region of the main density, may still easily detect the
halo structure.  Then, besides the possible halo related to the $2p_{1/2}$
level, the interesting region of possible one-neutron halo heavier than 
$^{37}$Mg will be the neutron-drip-line nucleus with $N \approx 50$ coming from
the weakly-bound 3$s_{1/2}$ orbit 
if the shape is limited to be spherical.    In contrast, 
if we look for the prolately deformed region as a result of near-degeneracy of
the $1g_{9/2}$, $3s_{1/2}$ and $2d_{5/2}$ orbits in the spherical limit, 
a deformed
$s$-wave halo may be expected for the lowest-lying 
$\Omega^{\pi} = 1/2^{+}$ orbit of
the $N_{ho}$=4 major shell, which will lie around the Fermi level of 
neutron-drip-line nuclei with $N \approx 40$.  
See Fig. 9. 

The reason for the overwhelming dominance of prolate shape compared with oblate
shape except for very light nuclei has not really pinned down.  The
clear difference of the splitting of one-particle levels coming from high-j
orbits on the prolate side from on the oblate side may have something to do with
the dominance.  Due to the unique shell-structure on the oblate side 
the particle numbers, at which the
large energy gap occurs on the oblate side of the Nilsson diagram of realistic
potentials, can be reliably guessed from the magic numbers of the deformed 
oscillator potential.  
In this connection, it would be interesting to know whether the shape of
$^{20}_{6}$C$_{14}$ is oblate or not.

At the end, once more I would like to express my sincere and heartfelt thanks 
to Aage Bohr and Ben Mottelson for guiding me for years to study and enjoy 
physics.

\newpage

\begin{table*}[t]
\caption{\label{tab:table1} The decay scheme of the septuplet 
$(h_{9/2} \, 3^{-}) \, I^{\pi}$ in $^{209}$Bi.    In the second column 
the calculated and experimental  
energy shifts from the unperturbed energy 2.614 MeV are shown. 
 In the third, fourth and fifth
columns the values of $B(E \nu)$ and 
$B(M \nu)$ are shown in respective Weisskopf units, where  
the number without brackets is the calculated value and the one with brackets is
the experimental value taken from Ref. \cite{JWH69}.  The numbers in parentheses
express the branching ratios.  In the calculation the values
$[e^{p}_{eff}(E1)]^{2}$ = 0.14 $e^{2}$, $g(3^{-})$ = $0.58$, 
$g_{s,eff}(M \nu)$ = 0.35 $g_{s,free}$ and $e^{p}_{eff}(E3)$ = 1.3 $e$ were
used.   The calculated results are taken from Ref. \cite{IH74}.     }

\begin{tabular}{|c|c|c|c|c|} \hline

$I^{\pi}$ & $\delta E_{calc} / \delta E_{exp}$ &
$\rightarrow \, 9/2^{-} \, (g.s.)$ &
$\rightarrow \, 7/2^{-} \, (0.89 \, \mbox{MeV})$  & 
$\rightarrow \, 13/2^{+} \, (1.60 \, \mbox{MeV})$ \\   
 & (keV) & & &  \\ \hline

3/2$^{+}$ & $-$190/$-$120 & 20.8 E3 (99) & 3.8 x 10$^{-2}$ M2 (1) &  \\
 &  & [(100)] &  &  \\ \hline   
9/2$^{+}$ & $-$89/$-$49 & 1.1 x 10$^{-3}$ E1 (99) & 3.1 x 10$^{-5}$ E1 (1) &  \\
 &  & [(1.4$\pm$1.8) x 10$^{-3}$ E1 (100)] & [$\leq$ (2.5$\pm$3) x 10$^{-4}$
 E1 ($\leq$ 5)] &   \\ \hline 
7/2$^{+}$ & $-$6/$-$29 & 1.5 x 10$^{-5}$ E1 (30) & 1.2 x 10$^{-4}$ E1 (70) &  \\
 &  & [(1.4$\pm$0.8) x 10$^{-5}$ E1 (33)] &  [(1$\pm$0.5) x 10$^{-4}$ E1 (67)] 
 &  \\ \hline 
11/2$^{+}$ & $-$31/$-$14 & 6.1 x 10$^{-4}$ E1 (99.5) & & 3.6 x 10$^{-3}$ M1 (0.5)\\
 & & [(1.7$\pm$1) x 10$^{-4}$ E1 (85)] & 
 & [(6.2$\pm$4) x 10$^{-2}$ M1 (15)]  \\ \hline
13/2$^{+}$ & $-$63/$-$14 & 30.2 E3 + 3.5 x 10$^{-3}$ M2 (14) & & 6.7 x 10$^{-3}$ M1
(86) \\  
 & & [(1)] & & [(1.1$\pm$0.6) x 10$^{-1}$ M1 (99)]  \\ \hline  
5/2$^{+}$ & +7/+4 & 31.9 E3 + 8.6 x 10$^{-3}$ M2 (49) & 2.2 x 10$^{-6}$ E1
(51)  \\
 & & [(41)] & [(2.2$\pm$1) x 10$^{-6}$ E1 (59)] \\ \hline
15/2$^{+}$ & +156/+130 & 30.3 E3 (55) & & 9.3 x 10$^{-4}$ M1 (45) \\
 & & [(53)] & & [(7$\pm$2) x 10$^{-4}$ M1 (47)] \\
\hline

\end{tabular}

\end{table*}


\begin{table*}[ht]
\caption{\label{table.2}  Probability for bound one-neutron wave-functions
to remain inside a finite square-well potential with radius $R_{0}$, in the
limit that 
eigenenergies $\varepsilon_{n \ell}$ ($< 0$) approaches zero.      }

\begin{tabular}{|c|c|c|c|c|c|} \hline

$\ell$ & 0 & 1 & 2 & 3 & $\ell$ ($\neq$ 0)   \\ \hline
$\int^{R_{0}}_{0} \mid R_{n \ell}(r) \mid^{2}$ dr & 0 & 1/3 & 3/5 & 5/7 & 
($2\ell - 1$)/($2\ell + 1$)  \\

\hline

\end{tabular}

\end{table*}


\begin{figure*} [ht]
\includegraphics[width=\textwidth]{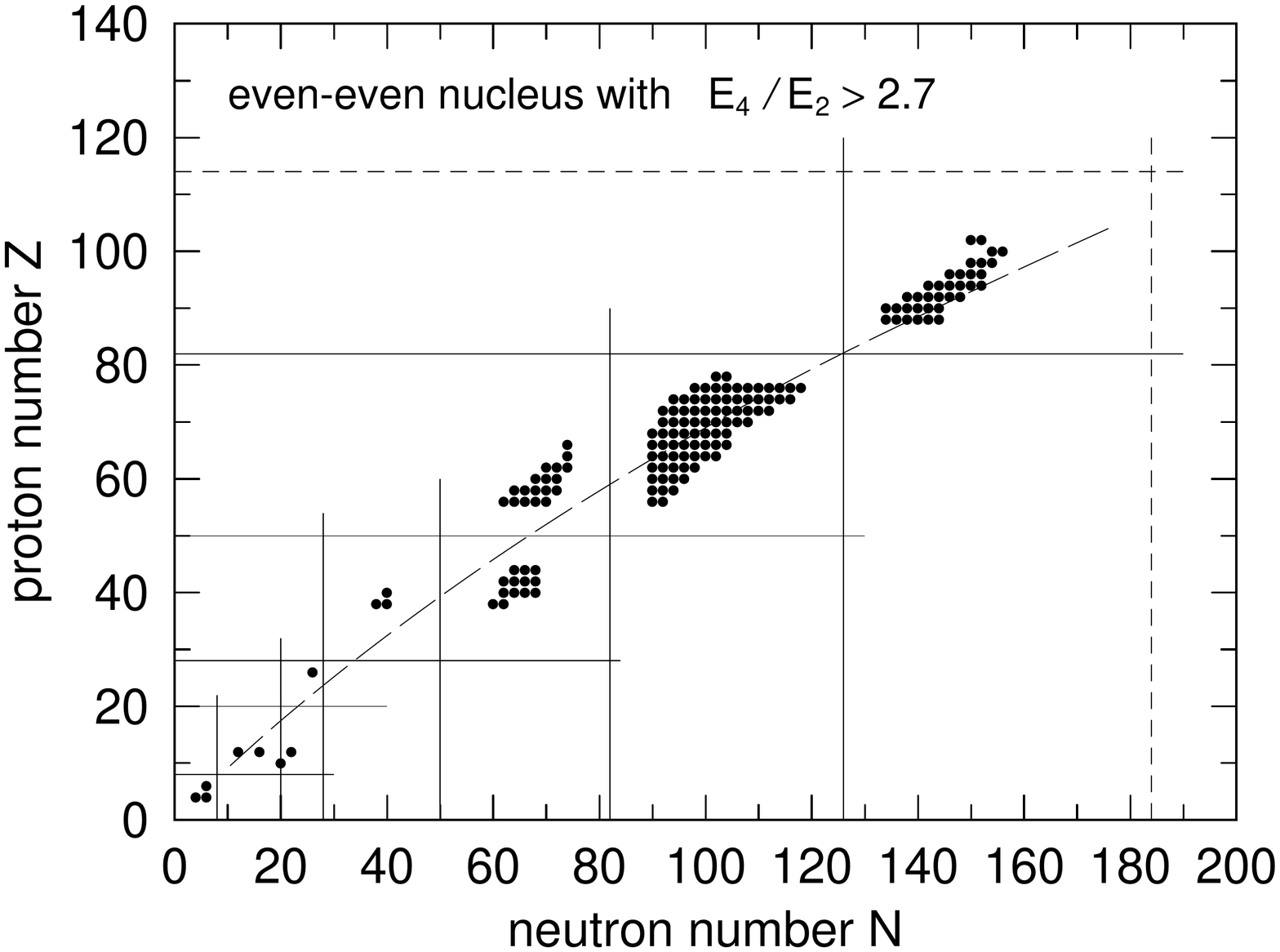}
\caption{
Regions of deformed even-even nuclei. Even-even nuclei have the ground-state
spin-parity $0^{+}$, without exception.  The overwhelming majority of these has
$2^{+}$ first excited state.  Writing the excitation energies of the
lowest-lying $2^{+}$ and $4^{+}$ states as $E(2_{1}^{+})$ and $E(4_{1}^{+})$, 
the filled circle denotes even-even nuclei, in which 
$E(4_{1}^{+})/E(2_{1}^{+}) > 2.7$ .  The data are taken from
http://www.nndc.bnl.gov/ensdf/.   
The line of $\beta$-stability is indicated by the thin long-dashed curve. 
The thin straight lines parallel to the x and y axes show the magic numbers of
protons and neutrons, which are known in nuclei along the $\beta$-stability
line.  
Except for very light nuclei (Z$\leq$8) the
neutron drip line, at which nuclei become unstable for neutron emission, is not
known experimentally.  The border of deformed nuclei plotted for the neutron-rich
region of medium-heavy nuclei is often equal to the border of neutron-rich
nuclei, for which the energy of the 4$_{1}^{+}$ state is presently known.  
}
\end{figure*}


\begin{figure*} [ht]
\includegraphics[width=\textwidth]{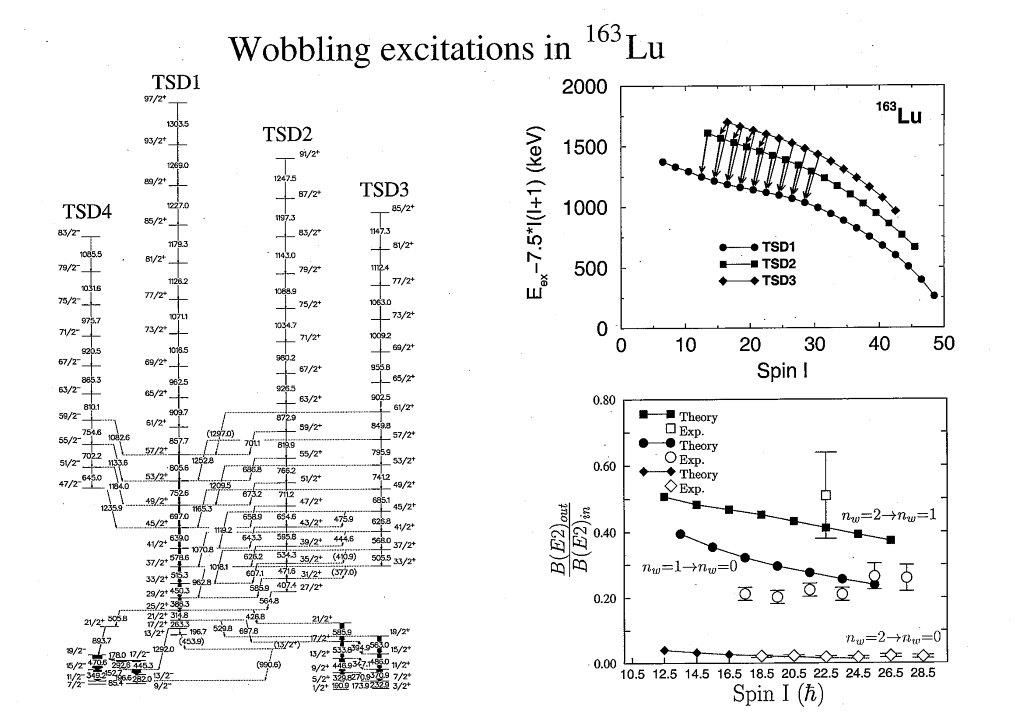}
\caption{
Experimental data on the wobbling excitations in $^{163}$Lu.  
The figures on the left and the upper right are made from measured level scheme,
while measured B(E2) values in comparison
with calculated B(E2) values are shown in the figure on the lower right.    
B(E2)$_{in}$ expresses B(E2; $n_{W}$, I 
$\rightarrow$ $n_{W}$, I$-$2), while B(E2)$_{out}$ denotes 
B(E2; $n_{W}$, I $\rightarrow$ $n_{W}-1$, I$-$1) or B(E2; $n_{W}$, I
$\rightarrow$ $n_{W}-2$, I$-$2).  TSD2 is identified as the
one-phonon ($n_{W}$=1) wobbling band built on the yrast TSD1 ($n_{W}$=0), while
TSD3 as the two-phonon ($n_{W}$=2) wobbling band.  
The author expresses her thanks to G. B. Hagemann for the present figure.  
}
\end{figure*}


\begin{figure*} [ht]
\includegraphics[width=0.6\textwidth]{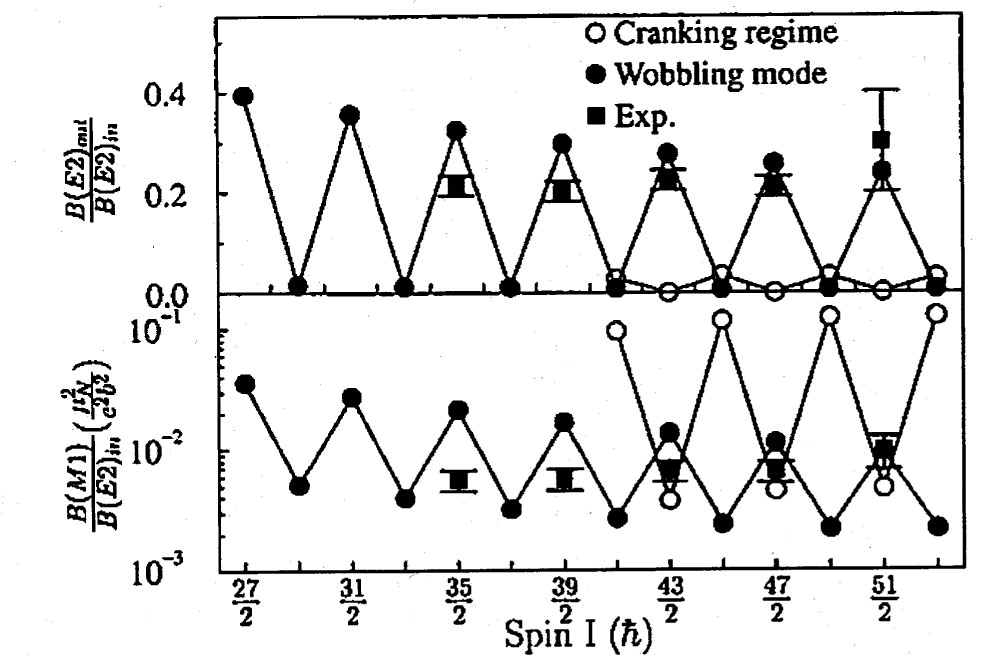}
\caption{
Experimental electromagnetic properties of the transitions
connecting TSD2 with TSD1 are shown by filled squares with experimental errors. 
B(E2)$_{out}$
 represents B(E2; TSD2, I $\rightarrow$ TSD1, I$-$1), and 
 B(M1) expresses B(M1; TSD2, I $\rightarrow$ TSD1, I$-$1), while 
 B(E2)$_{in}$ denotes B(E2; TSD2, I $\rightarrow$ TSD2, I$-$2). 
 Filled circles denote calculated values obtained from the particle-rotor model,
 of which the result shows that the wobbling excitation becomes the lowest
 unfavored-signature ($\alpha_{u}$) state 
 in the relevant angular-momentum region, while open circles
 represent calculated values obtained by using cranking model.   
The figure is taken from Ref. \cite{SWO01}}
\end{figure*}


\begin{figure*} [ht]
\includegraphics[width=0.6\textwidth]{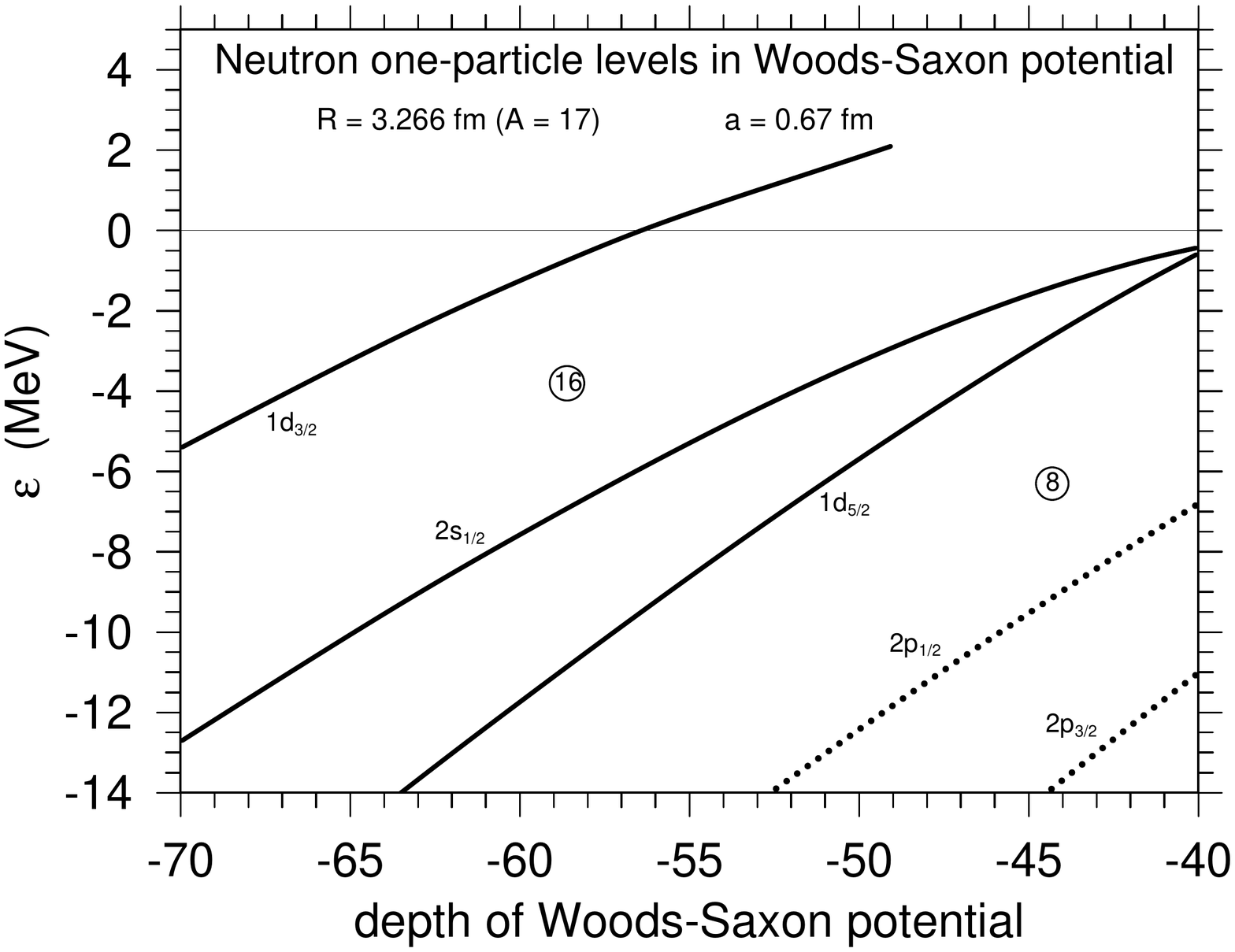}
\caption{
Calculated neutron one-particle energies as a function of the depth of spherical
Woods-Saxon potentials.  The $\ell$=2 one-particle resonant level continues to be
well defined up to $\varepsilon$ = 2 MeV, while there is no one-particle
resonance for $\ell$=0.  
The parameters of the Woods-Saxon potential except for the depth are kept
constant and are designed
approximately for the nucleus $^{17}$C, for which the realistic  
depth should be
around $-$40 MeV. 
The figure is taken from Ref. \cite{IH07}.
}
\end{figure*}


\begin{figure*} [ht]
\includegraphics[width=\textwidth]{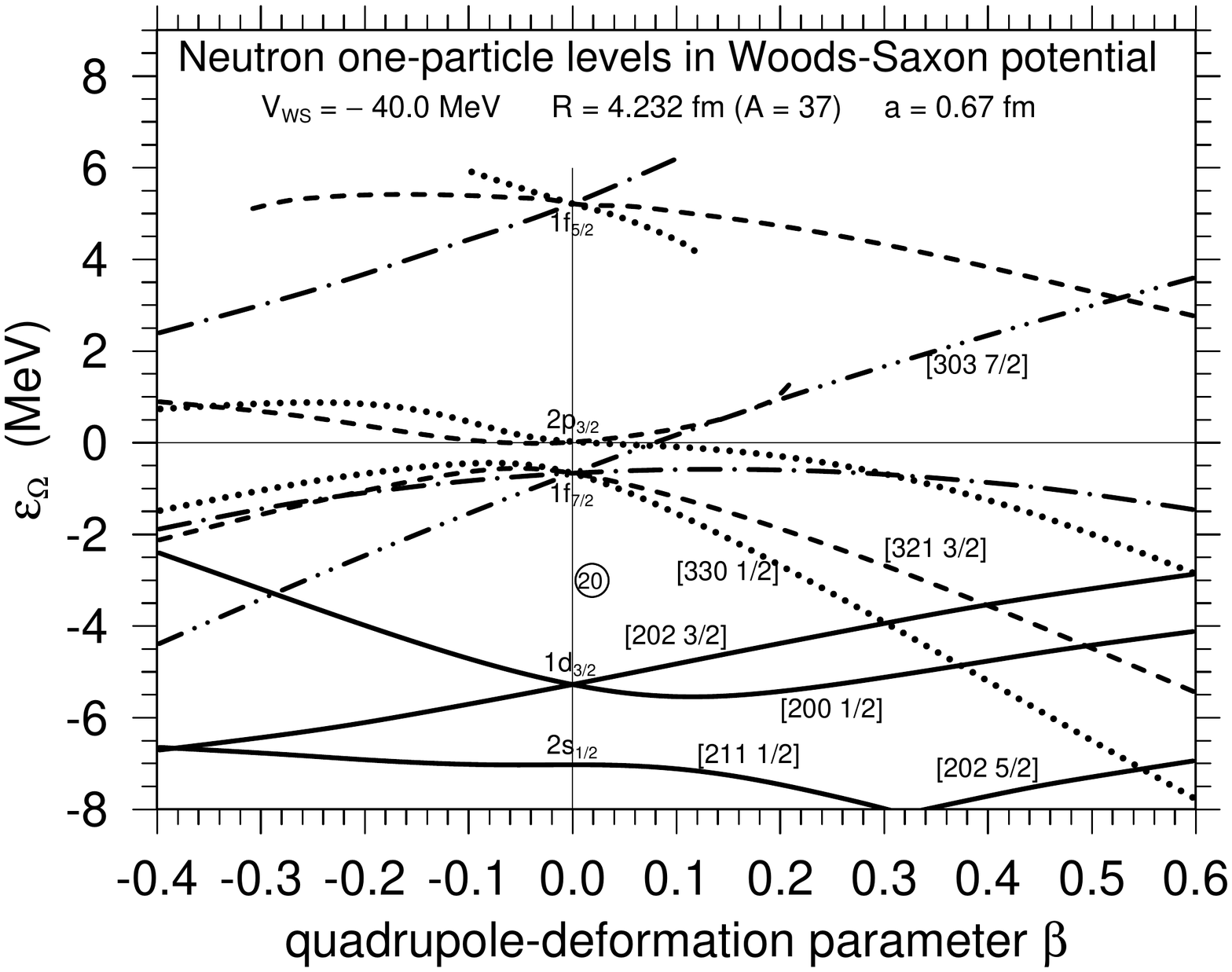}
\caption{
Calculated neutron one-particle levels as a function of 
axially-symmetric quadrupole
deformation.  Parameters of the Woods-Saxon potential are designed approximately
for the nucleus $^{37}$Mg.  
Note that the $2p_{3/2}$ level at $\beta$=0 is a one-particle resonant level
with the energy $+$0.018 MeV.
The $2p_{1/2}$ resonant level is not obtained at
$\beta$=0, and for $\beta \neq 0$ no $\Omega^{\pi}$ = 1/2$^{-}$ 
one-particle level
connected to the possible $2p_{1/2}$ level can survive as a resonant level.  
The $\Omega^{\pi}$ = $1/2^{-}$ levels are denoted by dotted curves, the
$\Omega^{\pi}$ = $3/2^{-}$ levels by dashed curves, the 
$\Omega^{\pi}$ = $5/2^{-}$ levels by dot-dashed curves, and 
the $\Omega^{\pi}$ = $7/2^{-}$ levels by dot-dot-dashed curves, while
positive-parity levels are plotted by solid curves.  
The
$\Omega^{\pi}$ = 1/2$^{-}$ resonant level 
connected to the one-particle resonant $1f_{5/2}$ level at $\beta$=0 
cannot survive as a resonance for 
$\beta > 0.12$ because of the increasing $\ell$=1 component inside the nuclear
radius.  The definition of one-particle resonance in a deformed potential can be
found in \cite{RGN66, IH05}.  
The figure is taken from Ref. \cite{IH07}.
}
\end{figure*}



\begin{figure*} [ht]
\includegraphics[width=\textwidth]{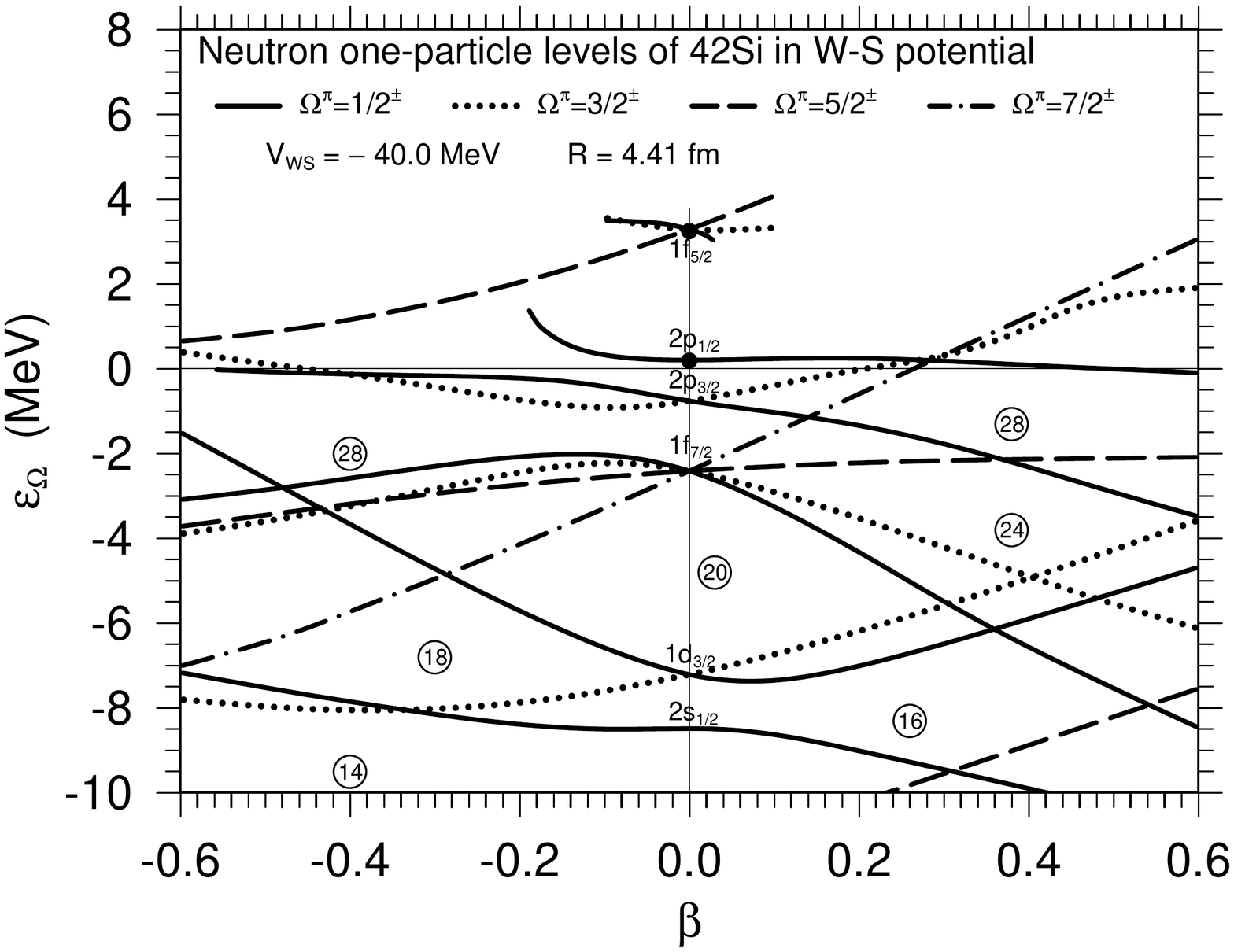}
\caption{
Calculated one-particle energies for neutrons of $^{42}_{14}$Si$_{28}$ as a
function of axially-symmetric quadrupole deformation.  Bound one-particle
energies at $\beta$=0 are $-8.50$, $-7.24$, $-2.42$ and $-0.77$ MeV for the
$2s_{1/2}$, $1d_{3/2}$, $1f_{7/2}$ and $2p_{3/2}$ levels, respectively, while
one-particle resonant $2p_{1/2}$ and $1f_{5/2}$ levels are obtained at +0.20 MeV
with the width of 0.20 MeV and +3.25 MeV with the width of 0.58 MeV,
respectively, which are denoted by filled circles.  One-particle resonant
energies for $\beta \neq 0$ are not plotted unless they are important for the
present discussion.  For simplicity, calculated widths of one-particle resonant
levels are not shown.  The neutron numbers, which are obtained by filling all
lower-lying levels, are indicated with circles.  The parity of levels can be
seen from the $\ell$ values denoted at $\beta$=0; $\pi = (-1)^{\ell}$.
}
\end{figure*}



\begin{figure*} [ht]
\includegraphics[width=\textwidth]{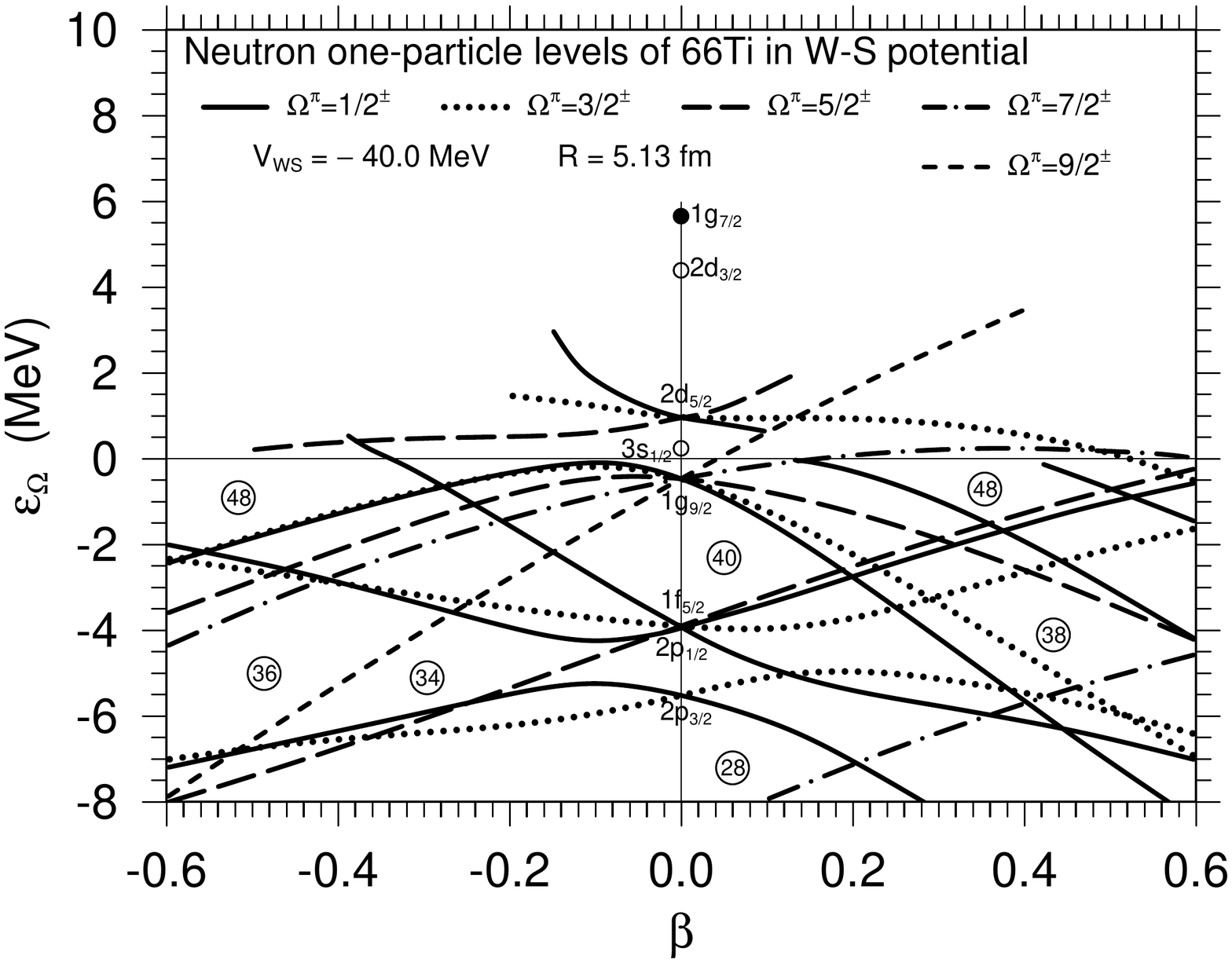}
\caption{
Calculated neutron one-particle energies as a function of axially-symmetric
quadrupole deformation.
Parameters of the Woods-Saxon potential are chosen for the nucleus 
$^{66}_{22}$Ti$_{44}$.  Bound one-particle energies at $\beta = 0$ are $-8.82$, 
$-5.54$, $-3.99$, $-3.94$, and $-0.48$ MeV for the $1f_{7/2}$, $2p_{3/2}$,
$2p_{1/2}$, $1f_{5/2}$, and $1g_{9/2}$ levels, respectively, while one-particle
resonant $2d_{5/2}$, $1g_{7/2}$, and $1h_{11/2}$ levels are obtained at $+$0.96,
$+$5.66, and $+$7.57 MeV, respectively.  The $2d_{3/2}$ one-particle resonant
level is not obtained for the present potential, however, its approximate
position at $\beta = 0$ is denoted by an open circle, at which an eigenphase
does not reach, but comes close to $\pi /2$.  The $3s_{1/2}$ resonant level does
not exist in any case, but the open circle at $\beta = 0$ indicates the energy
obtained by extrapolating the solid curve of the bound 
$\Omega^{\pi}$ = 1/2$^{+}$ orbit for $\beta > 0.12$ to $\beta = 0$, although the
calculated solid curve reaches 0 at $\beta = 0.12$ and cannot continue to $\beta
< 0.12$.  The major component of the solid curve for 
$\varepsilon_{\Omega} (<0) \rightarrow 0$ is clearly $3s_{1/2}$.  One-particle
resonant levels for $\beta \neq 0$ are not plotted if they are not relevant for
the present discussion.  The neutron numbers, 28, 34, 36, 38, 40 and 48, which
are obtained by filling all lower-lying levels, are indicated with open circles.
  One-particle levels with $\Omega$ = 1/2, 3/2, 5/2, 7/2 and 9/2 are expressed
  by solid, dotted, long-dashed, dot-dashed, and short-dashed curves,
  respectively, for both positive and negative parities. 
}
\end{figure*}



\vspace{2cm}
\begin{thebibliography}{99}
\bibitem{BM75} A. Bohr and B. R. Mottelson, {\it Nuclear Structure\/} (Benjamin,
Reading, MA, 1975), Vol. II.  
\bibitem{IH74} I. Hamamoto, Physics Reports {\bf 10C}, no.2, (1974). 
\bibitem{BP69} A. R. Barnett and W. R. Phillips, Phys. Rev. {\bf 186}, 1205
(1969). 
\bibitem{HKW72} O.Hausser, F. C. Khanna and D. Ward, Nucl. Phys.  {\bf A194},
113 (1972).
\bibitem{GDH71} E. Grosse {\it et al.}, Nucl. Phys. {\bf A174}, 525, (1971). 
\bibitem{JWH69} J. W. Hertel, D. G. Fleming, J. P. Schiffer and H. E. Gove,
Phys. Rev. Lett. {\bf 23}, 488 (1969). 
\bibitem{IH77} I. Hamamoto, Phys. Lett. {\bf 66B}, 410, (1977).
\bibitem{MN72} M. Nagao, in: Proc. of the Intern. Conf. on Nuclear Structure
Studies using Electron Scattering and Photoreaction, eds. K. Shoda and H. Ui
(Sendai, 1972) p.121.
\bibitem{JH70} J. H. Heisenberg and I. Sick, Phys. Lett. {\bf 32B}, 249 (1970).
\bibitem{MP82} M. Piiparinen {\it et al.}, Z. Phys.  {\bf A309}, 87, (1982).
\bibitem{PK82} P. Kleinheinz, {\it et al.}, Phys. Rev. Lett. {\bf 48}, 1457, 
(1982).
\bibitem{SGN55} S. G. Nilsson, Mat. Fys. Medd. Dan. Vid. Selsk.  {\bf 29}, 
no.16, (1955).
\bibitem{SAM55} S. A. Moszkowski, Phys. Rev. {\bf 99}, 803, (1955).
\bibitem{KG56} K. Gottfried, Phys. Rev. {\bf 103}, 1017, (1956).
\bibitem{AJ71} A. Johnson, H. Ryde and J. Sztarkier,  Phys. Lett. {\bf B34}, 
605, (1971); A. Johnson, H. Ryde and S. A. Hjorth, Nucl. Phys. {\bf A179}, 753
(1972).
\bibitem{CPS74} S. Cohen, F. Plasil and W. J. Swiatecki, Ann. Phys. {\bf 82},
557 (1974).
\bibitem{AB75} A. Bohr, Nobel Lecture, Dec.11, 1975 (Nobelstiftelsen 1976).
\bibitem{BM79} A. Bohr and B. R. Mottelson, Phys. Today, {\bf 32}, 25 (1979).  
\bibitem{IH85} I. Hamamoto, Treatise on Heavy-Ion Science, Vol.3, 313,
(Plenum Publishing Corporation, 1985).
\bibitem{GHH86} J. D. Garrett, G. B. Hagemann and B. Herskind, Ann. Rev. Nucl.
Part. Sci. {\bf 36}, 419 (1986).  
\bibitem{BM74} A. Bohr and B. R. Mottelson, Phys. Scri. {\bf 10A}, 13 (1974).
\bibitem{BHM78} R. Bengtsson, I. Hamamoto and B. R. Mottelson, Phys. Lett. {\bf
73B}, 259 (1978).
\bibitem{BF79} R. Bengtsson and S. Frauendorf, Nucl. Phys. {\bf A327}, 139 
(1979).
\bibitem{IH85BM} I. Hamamoto, NUCLEAR STRUCTURE 1985, p.129, Proc. of the Niels
Bohr Centennial Conference, Copenhagen, May, 1985 (NORTH-HOLLAND).  
\bibitem{IH76} I. Hamamoto, Nucl. Phys. {\bf A271}, 15, (1976).
\bibitem{PT86} P. Twin {\it et al.},   Phys. Rev. Lett. {\bf 57}, 811 (1986). 
\bibitem{SMP62} S. M. Polikanov {\it et al.}, J. Exptl. Theoret. Phys. (USSR)
{\bf 42}, 1464 (1962); Transl. Soviet Phys. JETP {\bf 15}, 1016 (1962).  
\bibitem{NT88} P. J. Nolan and P. J. Twin, Ann. Rev. Nucl. Part. Sci. {\bf 38}, 
533 (1988).  
\bibitem{JK91} R. V. F. Janssens and T. L. Khoo, Ann. Rev. Nucl. Part. Sci. {\bf
41}, 321 (1991).
\bibitem{HW11} Some experimental data are collected in: K. Heyde and J. L. Wood,
 Rev. Mod. Phys. {\bf 83}, 1467 (2011).
\bibitem{IH90} For example, see: I.Hamamoto, Nucl. Phys. {\bf A520}, 297c 
(1990). 
\bibitem{SWO01} S.W. \O deg\aa rd {\it et al.}, Phys. Rev. Lett. {\bf 86}, 5866
(2001).  
\bibitem{DRJ02} D. R. Jensen {\it et al.}, Phys. Rev. Lett. {\bf 89}, 142503
(2002).    
\bibitem{IH02} I. Hamamoto, Phys. Rev. C {\bf 65}, 044305 (2002).  
\bibitem{HM83} I. Hamamoto and B. R. Mottelson, Phys. Lett. {\bf B127}, 281
(1983).  
\bibitem{GA76} G. Andersson {\it et al.}, Nucl. Phys. {\bf A268}, 205 (1976).  
\bibitem{IH87} I. Hamamoto, Phys. Lett. {\bf B193}, 399 (1987). 
\bibitem{DHH02} D. R. Jensen {\it et al.}, Nucl. Phys. {\bf A703}, 3 (2002). 
\bibitem{GBH04} G. B. Hagemann, private communications.  
\bibitem{HSP95} H. Schnack-Petersen {\it et al.}, Nucl. Phys. {\bf A594}, 175
(1995).  
\bibitem{SF97} S. Frauendorf and J. Meng, Nucl. Phys. {\bf A617}, 131 (1997).
\bibitem{KS01} K. Starosta {\it et al.}, Phys. Rev. Lett. {\bf 86}, 971 (2001).
\bibitem{TK04} T. Koike, K. Starosta and I. Hamamoto, Phys. Rev. Lett. {\bf 93},
172502 (2004).  
\bibitem{BM69} A. Bohr and B. R. Mottelson, {\it Nuclear Structure\/} (Benjamin,
Reading, MA, 1969), Vol. I.  
\bibitem{AO00} A. Ozawa, T. Kobayashi, T. Suzuki, K. Yoshida, and I. Tanihata, 
 Phys. Rev. Lett. {\bf 84}, 5493 (2000).
\bibitem{JT37} H. Jahn and E. Teller, Proc. R. Soc. London, Ser. {\bf A161}, 
220 (1937). 
\bibitem{IH07} I. Hamamoto, Phys. Rev.  {\bf C76}, 054319 (2007). 
\bibitem{IH12} I. Hamamoto, Phys. Rev.  {\bf C85}, 064329 (2012).  
\bibitem{RGN66} R. G. Newton, {\it Scattering Theory of Waves and Particles} 
(McGraw-Hill, New York, 1966).  
\bibitem{IH05}  I. Hamamoto, Phys. Rev. {\bf C72}, 024301 (2005); {\bf C73},
064308 (2006).  
\bibitem{PD13} P. Doornenbal {\it et al.}, Phys. Rev. Lett. {\bf 111}, 212502 
(2013).  
\bibitem{TN09} T. Nakamura {\it et al.}, Phys. Rev. Lett. {\bf 103}, 262501 
(2009); {\bf 112}, 142501 (2014).  
\bibitem{NK14} N. Kobayashi {\it et al.}, Phys. Rev. Lett. {\bf 112}, 242501 
(2014).
\bibitem{IT85} I. Tanihata {\it et al.}, Phys. Rev. Lett. {\bf 55}, 2676 (1985).
\bibitem{PF10} P. Fallon {\it et al.}, Phys. Rev. {\bf C81}, 041302 (2010).
\bibitem{HM03} I. Hamamoto and B. R. Mottelson, C. R. Physique {\bf 4}, 433 
(2003).  
\bibitem{TM97} T. Misu, W. Nazarewicz and S. \AA berg, Nucl. Phys. {\bf A614}, 44 
(1997).  
\bibitem{IH04} I. Hamamoto, Phys. Rev. {\bf C69}, 041306(R) (2004).
\bibitem{ST12} S. Takeuchi {\it et al.}, Phys. Rev. Lett. {\bf 109}, 182501 
(2012).
\bibitem{IH14} I. Hamamoto, Phys. Rev. {\bf C89}, 057301 (2014).
\bibitem{HM09} I. Hamamoto and B. R. Mottelson, Phys. Rev. {\bf C79}, 034317
(2009).
\bibitem{MV60} B. R. Mottelson and J. G. Valatin, Phys. Rev. Lett. {\bf 5}, 511
(1960).









\end{thebibliography}
\end{document}